\documentclass[conference,twoside]{IEEEtran}
\usepackage{graphicx}
\usepackage{amsmath}
\usepackage{amsfonts}
\usepackage{amsthm}
\usepackage[mathscr]{euscript}
\usepackage{mdwmath}
\usepackage{bbm}
\usepackage{mathrsfs}
\usepackage{bm}
\usepackage{epstopdf}
\usepackage{booktabs,fixltx2e}
\usepackage[flushleft]{threeparttable}
\usepackage{siunitx}
\graphicspath{{./figures/}}
\usepackage[tight,footnotesize]{subfigure}
\usepackage{cite}
\usepackage{wrapfig}
\usepackage{algorithm}
\theoremstyle{plain}

\usepackage[noend]{algpseudocode}
\usepackage{float}

\usepackage{comment}
\usepackage{caption}

\usepackage{marginnote}
\usepackage{color}
\usepackage{tabularx}
\usepackage{multicol,lipsum}

\usepackage{algorithm}
\usepackage[]{algpseudocode}

\newcommand{\mydef}{:=}
\newcommand{\dx}{\dot{x}}
\newcommand{\Rc}{\mathbb{R}}
\newcommand{\omegas}{\omega_s}

\newcommand{\domg}{\dot{\omega}}
\newcommand{\ddel}{\dot{\delta}}
\newcommand{\hatomg}{\hat{\omega}}
\newcommand{\hatdel}{\hat{\delta}}
\newcommand{\ocoi}{\omega_{COI}}
\newcommand{\dcoi}{\delta_{COI}}

\allowdisplaybreaks

\renewcommand{\baselinestretch}{1}
\usepackage[left=1.91cm, right=1.91cm, top=2.54cm, bottom=1.91cm]{geometry}

\begin{document}

\title{Influence of Load Models on Equilibria, Stability and Algebraic Manifolds of Power System Differential-Algebraic System}
\author{
\IEEEauthorblockN{Dan Wu and Bin Wang}
	
}

\maketitle
\begin{abstract}
Load models have a great impact on voltage behaviors as well as power system transient dynamics. Extensive work has been done on this topic, proposing appropriate load models and capturing better load behaviors during transient. This paper presents a comprehensive study to investigate the geometric and topological changes induced by different load models for the traditional power system differential-algebraic equations. Specifically, we attempt to reveal the deformation of equilibria, stability regions, and algebraic manifolds during a continuous evolution of load model. Several findings are presented in the paper, some of which countering traditional recognitions and intuitions. A major discovery is that the load model with a large proportion of constant impedance and a small proportion of constant power exhibits much more complex features than the load model with the reversed proportions of impedance and power. The increase of complexity is thoroughly recorded and investigated by the changes of geometric properties and mutations of topological invariants in the sense of equilibria, stability regions, and algebraic manifolds for the DAE system. However, most of the changes seem to occur on unstable components of algebraic manifolds or near the singular boundary surfaces, suggesting a limited variation of dynamical behaviors on the stable component. 
\end{abstract}


\section{Introduction}
   \label{sec:intro}
   Load behaviors have a profound impact on power system dynamical performance. Inaccurate load models can introduce significant errors in stability analysis \cite{lee1987:load4stability}, fail to capture important phenomena, e.g. fault-induced delayed voltage recovery \cite{Kosterev2008,Kosterev2009:presentation}, and even lead to power systems operated in modes where collapse and separation may occur \cite{cigre1990}. 

Power system engineers and researchers have been continuously improving the accuracy of load models in the past several decades \cite{taskforce1993,taskforce1995,review2017,huang2017:clm}. These efforts along with advanced parameter estimation techniques significantly reduce the gap between the actual measured load behaviors and the numerically simulated behaviors based on the model. Introducing subtler load models can potentially better approximate the measured load behaviors. With today's computing capabilities, there seems to be no reason to avoid an accurate load model, if possible, in transient stability simulations. The state-of-the-art load model is the composite load model developed by the Western Electricity Coordinating Council, which includes static components, four motors and an electronic component \cite{review2017}. Further improvements are ongoing in the field \cite{huang2017:clm}.


Although multiple load models are available, most of them favor numerical simulations in a qualitative way. When applied to the analytical transient stability analysis such as direct methods, they would face several difficulties in either derivation or analysis, for example, formulating an appropriate transient energy function \cite{Min2007} and handling of differential-algebraic equations (DAEs) instead of ordinary differential equations (ODEs) \cite{padiyar2017}.
This is the reason that most analytical methods were developed on classical power system model whose loads are represented by constant impedance \cite{pai1981:stability,chiang2011:bcu}. Further generalizations are possible but not always easy \cite{hiskens1996,Min2007}. It has been reported that static load models with properly identified parameters are adequate for transient stability analysis, since the transient stability is mainly about the real power behavior of the load while the static load models can capture the real power behavior with a fairly acceptable accuracy \cite{chiang2007}. 

An earlier attempt in \cite{wu2006:geometry} studied the influence of load models on a power system DAE model using a simple system with a single machine, an infinite bus and one load. To study more complex behaviors, higher dimensional state spaces need to be investigated. Thus, this paper studies 5-Bus, 9-Bus, and 14-Bus test cases in the center-of-inertia framework, and presents extensive investigations on equilibria, stability regions and algebraic manifolds of differential-algebraic equations. Several new interesting phenomena are revealed, visualized and analyzed when loads are represented by a combination of constant power and constant impedance. At a fixed loading level, when loads transition gradually from the constant power model to the constant impedance model,
\begin{enumerate}
	\item the number of power flow solutions drastically increases, but not monotonically;
	\item there is always a single stable equilibrium point (SEP) on the stable component of the algebraic manifold;
	\item neither a type-1 unstable equilibrium point (UEP) on the stable component of the algebraic manifold should admit low voltage at only one bus, nor a solution with low voltage at only one bus should be type-1 UEP;
	\item along with the gradual disappearance of singular surface, the stability region expands during the load transition, determined initially by the singularity boundary and eventually by the stable manifolds of type-1 UEP(s).
	\item the algebraic manifold is enlarged during the load transition, connected to its modulo $2 \pi$ replica, and eventually occupies the entire angle subspace;
	\item the topological invariants of the algebraic manifold mutate, from a sphere to a quotient space of a few tori;
	\item the change of topology for the algebraic manifold only occurs on the unstable components.
\end{enumerate}

The rest of the paper is organized as follows: Section II introduces the power system differential-algebraic equations for transient stability analysis. Section III, IV and V respectively present numerical investigations of equilibria, stability region and algebraic manifold of the IEEE 9-bus system considering loads gradually transitioning from constant power model to constant impedance model. Section VI draws conclusions and envisions the future work.

\section{Power System Differential-Algebraic Model for Transient Stability Analysis}
	\label{sec:DAE}
	This section presents the modeling of power systems used for this study which can consider different load models. We first introduce the power system model represented by DAEs using the center-of-inertia (COI) framework. Then, we model the load as a combination of constant power load and constant impedance. With an introduced parameter to linearly combine these two load models, we design a continuous transition from one to the other. To identify multiple equilibrium points for these DAEs, an equivalent power flow problem in the COI framework is formualted. Finally, we show how to determine the type of the identified equilibrium points.

\subsection{Classical Model in COI Framework}
Consider a general $N$-bus power grid with $N_g$ generator buses, including PV and slack buses, and $N_d$ PQ buses. Appearently, $N=N_g+N_d$. We adopt the classical DAE power injection model \cite{sauer2017:power} for the transient stability analysis throughout this paper. A general form is presented by
\begin{subequations}
	\begin{align}
	&\dx=f(x,y) \label{eq:differential} \\
	&0=g(x,y) \label{eq:algebraic}
	\end{align} \label{eq:DAE}%
\end{subequations}%
where $x \in \Rc^{2 N_g}$ is the differential state vector; $y \in \Rc^{2 N}$ is the algebraic state vector; $f:\Rc^{2 N} \to \Rc^{2 N_g}$; $g: \Rc^{2 N} \to \Rc^{2 N}$.

For each generator bus\footnote{The generator bus in this dynamical model is the generator internal bus. The generator terminal bus is, thereby, a PQ bus.}, a pair of differential equations captures the angular dynamics of the machine.
\begin{subequations}
	\begin{align}
	&\domg_i=\frac{\omegas}{2 H_i} (P_{m,i}-P_{e,i}) - D_i (\omega_i-\omegas) \label{eq:omega} \\
	&\ddel_i=\omega_i-\omegas \label{eq:delta}
	\end{align} \label{eq:PV_diff}%
\end{subequations}%
where subscript $i=1,2,\dots,N_g$ is the index of generator bus; $\omega_i$ is the rotor angular velocity; $\delta_i$ is the rotor angle; $\omegas$ is the constant synchronous speed; $H_i$ is the inertia constant; $P_{m,i}$ is the constant mechanical power injected into the generator; $P_{e,i}$ is the electrical power delivered from the generator which is defined shortly below; $D_i$ is the damping coefficient.

Each generator bus also induces two algebraic equations.
\begin{subequations}
	\begin{align}
	P_{e,i}=& V_{i} \sum_{k=1}^{N} V_{k} \big( G_{i,k} \cos(\delta_i-\delta_k) \nonumber \\
	& + B_{i,k} \sin(\delta_i-\delta_k) \big) \label{eq:Pe} \\
	V_i=&V_{m,i} \label{eq:V}
	\end{align} \label{eq:PV_alg}%
\end{subequations}%
where $G_{i,k}$ is the $(i,k)$'s entry of bus conductance matrix; $B_{i,k}$ is the $(i,k)$'s entry of bus susceptance matrix; $V_{m,i}$ is the constant bus voltage magnitude.

We substitute \eqref{eq:Pe} into \eqref{eq:omega} to eliminate $P_{e,i}$, reducing the DAE system to $\Rc^{2 N + N_g}$. One can also substitute \eqref{eq:V} in all other equations to further reduce the DAE system. However, later in the paper we will need these voltage equations to search for other equilibria. Hence, we leave them explicit in our DAE system.

For each PQ bus, only two power balance equations need to be specified.
\begin{subequations}
	\begin{align}
	0=&P_{d,j}+ V_{j} \sum_{k=1}^{N} V_{k} \big( G_{j,k} \cos(\delta_j-\delta_k) \nonumber\\
	& + B_{j,k} \sin(\delta_j-\delta_k) \big) \label{eq:Pd} \\
	0=&Q_{d,j}+V_{j} \sum_{k=1}^{N} V_{k} \big( G_{j,k} \sin(\delta_j-\delta_k) \nonumber\\
	& - B_{j,k} \cos(\delta_j-\delta_k) \big) \label{eq:Qd}
	\end{align} \label{eq:PQ_alg}%
\end{subequations}%
where subscript $j=N_g+1,N_g+2,\dots,N$ is the index of PQ bus; $P_{d,j}$ and $Q_{d,j}$ are respectively the active and reactive power loads.

Therefore, the overall DAE system includes \eqref{eq:omega}, \eqref{eq:delta}, \eqref{eq:V}, \eqref{eq:Pd}, and \eqref{eq:Qd}.

In this paper, we adopt the COI angle framework\footnote{Other angle reference choices include a particular infinite bus or an arbitrary PV bus.}. Define the COI angular velocity and angle.
\begin{subequations}
	\begin{align}
	&\ocoi \mydef \frac{1}{M} \sum_{i=1}^{N_g} M_i \omega_i \label{eq:omg_coi} \\
	&\dcoi \mydef \frac{1}{M} \sum_{i=1}^{N_g} M_i \delta_i \label{eq:del_coi}
	\end{align} \label{eq:COI}%
\end{subequations}%
where $M_i = 2 H_i/\omegas$ and $M = \sum M_i$.

Let's consider
\begin{subequations}
	\begin{align}
	&\hatomg_i \mydef \omega_i - \ocoi \label{eq:omg_new} \\
	&\hatdel_i \mydef \delta_i - \dcoi \label{eq:del_new} \\
	&\hatdel_j \mydef \delta_j - \dcoi
	\end{align} \label{eq:new_COI}%
\end{subequations}%

Substitute \eqref{eq:new_COI} into \eqref{eq:PV_diff}, \eqref{eq:PV_alg} and \eqref{eq:PQ_alg} we get
\begin{subequations}
	\begin{align}
	\dot{\hatomg}_i =& \dot{\omega}_i - \frac{1}{M} \sum_{n=1}^{N_g} M_n \dot{\omega}_n \label{eq:homg} \\
	\dot{\hatdel}_i =& \hatomg_i \label{eq:hdel}\\
	V_i = &V_{m,i} \label{eq:hV}\\
	0=&P_{d,j} + V_{j} \sum_{k=1}^{N} V_{k} \big( G_{j,k} \cos(\hatdel_j-\hatdel_k) \nonumber \\
	 & + B_{j,k} \sin(\hatdel_j-\hatdel_k) \big) \label{eq:hPd} \\
	0=&Q_{d,j} + V_{j} \sum_{k=1}^{N} V_{k} \big( G_{j,k} \sin(\hatdel_j-\hatdel_k) \nonumber \\
	 & - B_{j,k} \cos(\hatdel_j-\hatdel_k) \big) \label{eq:hQd}
	\end{align} \label{eq:DAE_COI}%
\end{subequations}%
Eqt.~\eqref{eq:DAE_COI} is the DAE system for which we will investigate the transient dynamics. 
Note that $\sum M_i \hatomg_i=0$, suggesting that \eqref{eq:DAE_COI} still has one degree of degeneracy which comes from \eqref{eq:hdel}. We will deal with this issue shortly below when using an equivalent power flow problem to find the equilibria.

\subsection{Load Modeling}
Equation.~\eqref{eq:PQ_alg} presents the power balance relation at each PQ bus\footnote{We choose the flow convention that injecting power is the positive direction.}. A typical load model is the constant power model, assuming that power injections $P_{d,j}$ and $Q_{d,j}$ at each PQ bus are constants. This model is very useful for static voltage stability analysis and induces the traditional power flow problem. During the transient, however, voltages at PQ buses fluctuate, which can alter the power consumption from their designated values. To better capture the change of power consumption, the ``ZIP" model is formulated in the following way.
\begin{subequations}
	\begin{align}
	&P_{d,j} = P_{0,j} + I_{p,j} V_j + G_{d,j} V_j^2 \label{eq:ZIP_P} \\
	&Q_{d,j} = Q_{0,j} + I_{d,j} V_j + B_{d,j} V_j^2 \label{eq:ZIP_Q}
	\end{align} \label{eq:ZIP}%
\end{subequations}%
where $P_{0,j}$ and $Q_{0,j}$ are the constant active and reactive power; $I_{p,j}$ and $I_{d,j}$ are the constant active and reactive current; $G_{d,j}$ is the load conductance; $B_{d,j}$ is the load susceptance.

In this paper, we ignore the constant current part in \eqref{eq:ZIP}, i.e. $I_{p,j}=I_{q,j}=0$, to acquire a uni-directional change of load model from the constant power to the constant impedance. Consider
\begin{equation}
	P_{d,j}+jQ_{d,j} = \alpha (P_{0,j}+jQ_{0,j}) + (1-\alpha) V_j^2/Z_{d,j}  \label{eq:ZP}
\end{equation}
where $\alpha \in [0,1]$; $Z_{d,j}$ is the constant impedance. 

We can track the change of transient dynamics by continuously changing $\alpha$ from $1$ to $0$, which gradually convert the constant power model to the constant impedance model. $P_{0,j}+jQ_{0,j}$ is fixed to be the designated load power consumption. When solving the traditional power flow high voltage solution $V_{0,j}$ associated with $P_{0,j}+jQ_{0,j}$, we obtain the corresponding impedance by
\begin{equation}
	Z_{d,j} = V_{0,j}^2/(P_{0,j}+jQ_{0,j})
\end{equation}

This setting ensures that the high voltage solution is unchanged during the change of load model.

\subsection{Equivalent Power Flow Problem for COI Framework}
To evaluate the transient dynamics of the system \eqref{eq:COI}, we need to obtain its equilibrium points. 
Setting $\dot{\hatdel}_i$ to zero in \eqref{eq:hdel} implies that the relative angular velocity $\hatomg_i$ vanishes, which further suggests that the true angular velocity $\omega_i$ converges to the COI angular velocity $\ocoi$ that does not necessarily comply with the synchronous speed $\omegas$. In this case, $\dot{\delta}_i$ in \eqref{eq:delta} does not vanish. Hence, all the generator angles keep changing, while their relative differences stabilize. Therefore, to solve the relative angle differences we subtract all COI angles from the first COI angle. Specifically, we define
\begin{subequations}
	\begin{align}
	\tilde{\omega}_i \mydef& \hatomg_i - \hatomg_1 = \omega_i-\omega_1 \label{eq:delative_omg}\\
	\tilde{\delta}_i \mydef& \hatdel_i - \hatdel_1 = \delta_i-\delta_1 \label{eq:delative_del}\\
	\tilde{\delta}_j \mydef& \hatdel_j - \hatdel_1 = \delta_j-\delta_1 \label{eq:delative_delr}
\end{align} \label{eq:relative}%
\end{subequations}%

Assuming that $D_i=D_j$ for every pair $i$ and $j$, the algebraic equations we are going to solve with respect to $\tilde{\delta}_i$ are
\begin{subequations}
	\begin{align}
	0 =& \frac{\omegas}{2 H_i} (P_{m,i}-P_{e,i})
	-\frac{\omegas}{2 H_1} (P_{m,1}-P_{e,1}) \label{eq:AEomg2} \\
	0 = &V_{m,i}-V_i \label{eq:AEV2}\\
	0=&P_{d,j} + V_{j} \sum_{k=1}^{N} V_{k} \big( G_{j,k} \cos(\tilde{\delta}_j-\tilde{\delta}_k) \nonumber \\
	& + B_{j,k} \sin(\tilde{\delta}_j-\tilde{\delta}_k) \big) \label{eq:AEPd2} \\
	0=&Q_{d,j} + V_{j} \sum_{k=1}^{N} V_{k} \big( G_{j,k} \sin(\tilde{\delta}_j-\tilde{\delta}_k) \nonumber \\
	& - B_{j,k} \cos(\tilde{\delta}_j-\tilde{\delta}_k) \big) \label{eq:AEQd2}
	\end{align} \label{eq:AE_COI2}%
\end{subequations}%

Equation.~\eqref{eq:AE_COI2} is the equivalent power flow problem that provides the equilibrium points to the DAE system \eqref{eq:DAE_COI}. It is similar to the traditional power flow problem except that every active power balance equation at generator bus subtracts from the active power balance equation at a particular generator bus. 

\subsection{Determining Stability and Type of Equilibrium}
Suppose $(x_0,y_0)$ is an equilibrium point to the DAE system \eqref{eq:DAE}. Let's consider the Jacobian matrix $J_x$ for the dynamic states $x$. Suppose $\frac{\partial g}{\partial y} |_{(x_0,y_0)}$ is nonsingular, then
\begin{equation}
	J_x \mydef \bigg[\frac{\partial f}{\partial x} - \frac{\partial f}{\partial y} \bigg(\frac{\partial g}{\partial y}\bigg)^{-1} \frac{\partial g}{\partial x}\bigg]_{(x_0,y_0)} \label{eq:Jx}
\end{equation}

Gathering the eigenvalues of $J_x$ in $\Lambda = \{\lambda_1,\dots,\lambda_{2 N_g}\}$. The point $(x_0,y_0)$ is said to be type-$k$ if there exists exactly $k$ entries of $\Lambda$ whose real parts are positive. If $J_x$ is hyperbolic and type-$0$, then we say $(x_0,y_0)$ is an SEP.

Note that \eqref{eq:DAE_COI} has one degree of degeneracy, the Jacobian matrix $J_x$ of dynamic states will always have a zero eigenvalue. Ignoring this zero eigenvalue, the type and stability of an equilibrium point can still be determined by the rest of the eigenvalues.

\subsection{Model Modification and Computational Procedure}
As discussed in the above subsection, the types of equilibria are evaluated from the differential part of the DAE system \eqref{eq:DAE_COI}. Therefore, the static power flow problem is not sufficient to determine the equilibria and their types. In this paper, we add generator classical model to each PV bus of the power flow problem, set the generator internal voltage bus as the new PV bus, and revise the generator terminal bus to PQ bus. The modified system is then a DAE system with more nodes and components. For example, the 9-bus system turns out to be a 12-bus system after modification since it includes three generator internal buses. We still call this system the 9-bus system since readers would be more familiar with the standard power flow system. The tested systems presented in this paper are all selected from the standard power flow systems, namely, 5-bus case\cite{salam1989:parallel,salam1989:parallel2}, 9-bus case, and 14-bus case\cite{UW:archive}, with modifications on the dynamical parts. The dynamical parameters are selected from the Appendices of \cite{anderson2008:power}, and are included in Appendix~I of this paper.

The computational procedure is given below.
\begin{enumerate}
	\item Choose a static power flow problem. Solve the high voltage solution by a standard solver. 
	\item Add generator internal bus to each generator bus of the static power flow problem. Compute the internal bus voltage based on the previously solved high voltage solution. Revise the generator terminal bus to PQ bus.
	\item Compute the equivalent load impedance at the high voltage solution. Choose the load model with given proportions of constant power and constant impedance.
	\item Based on the new DAE model, solve the corresponding equivalent power flow problem \eqref{eq:AE_COI2} for many equilibrium points.
	\item Evaluate the types and properties of these equilibrium points by \eqref{eq:Jx}.
\end{enumerate}
  
\section{Equilibria with Different Load Models}
	\label{sec:equilibria}
   	This section applies an efficient method \cite{wu2019:holomorphic,lesieutre2015:efficient} to solve \eqref{eq:AE_COI2} for multiple equilibra of the DAE model \eqref{eq:DAE_COI}. We gradually change the load model from constant power to constant impedance and observe the change of equilibra.
\subsection{Number of Equilibria}
\begin{figure}[!ht]
	\centering
	\subfigure[5-Bus Case]{\label{fig:case5}\includegraphics[width=0.9\columnwidth]{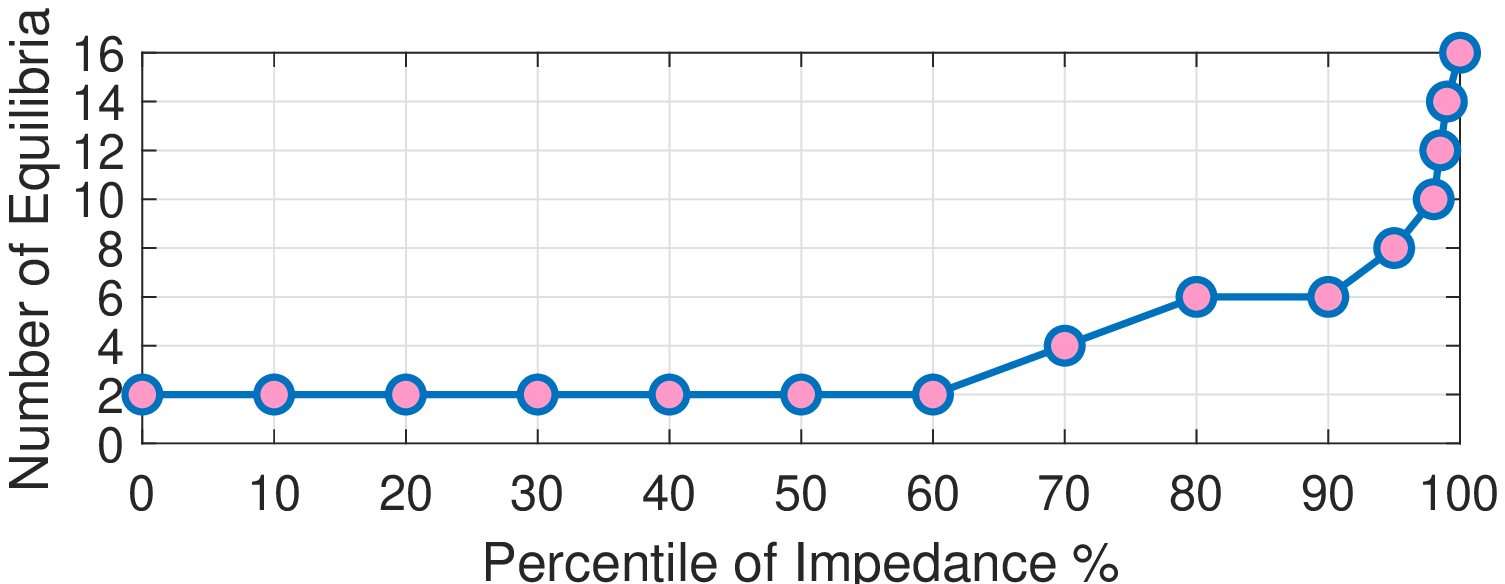}}
	\subfigure[9-Bus Case ]{\label{fig:case9}\includegraphics[width=0.9\columnwidth]{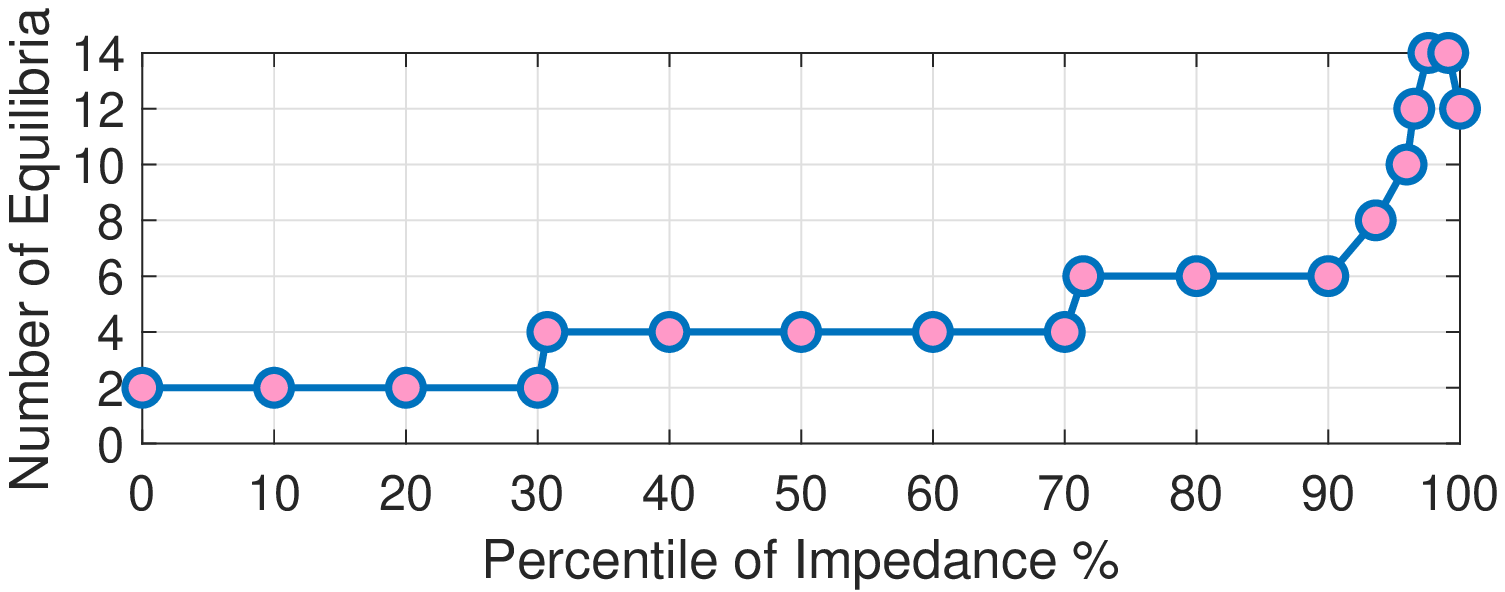}}
	\subfigure[14-Bus Case]{\label{fig:case14}\includegraphics[width=0.9\columnwidth]{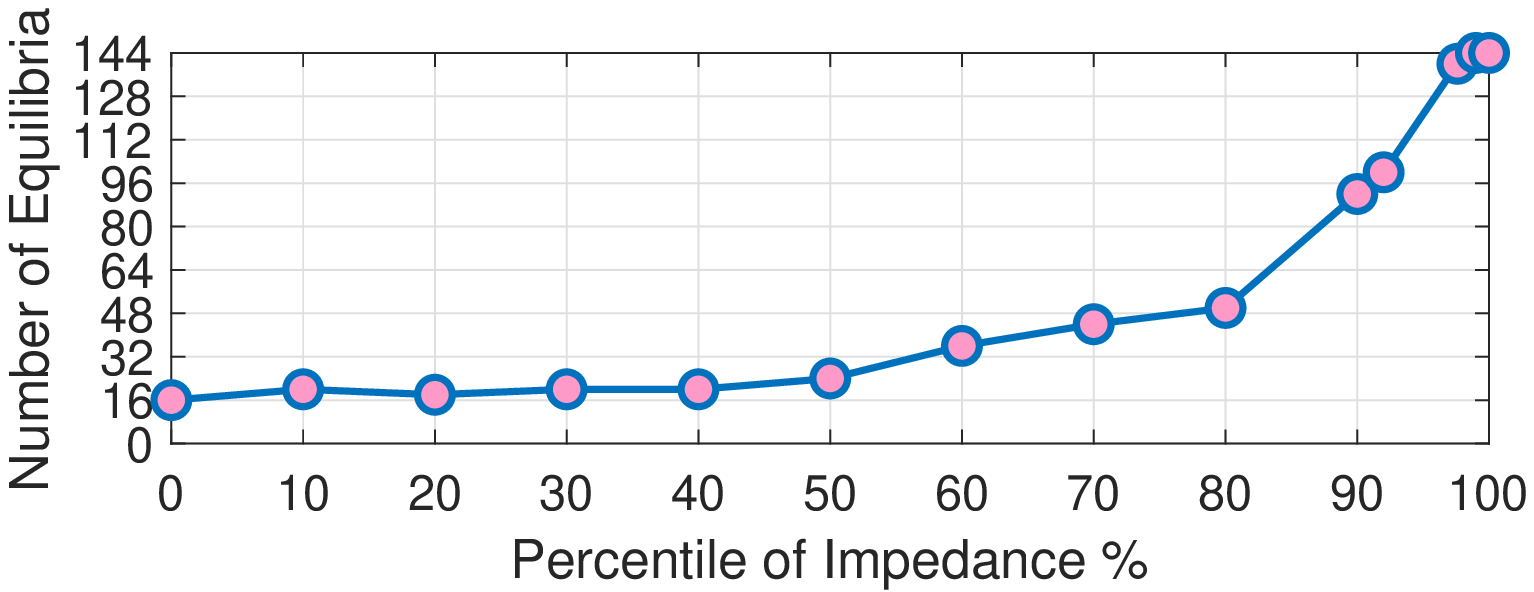}}
	\caption{Numbers of Equilibria with Different Load Models} \label{fig:num_equilibria}
\end{figure}
\begin{figure}[!ht]
	\centering
	\subfigure[Number of Equilibria Change]{\label{fig:case14zoom}\includegraphics[width=0.9\columnwidth]{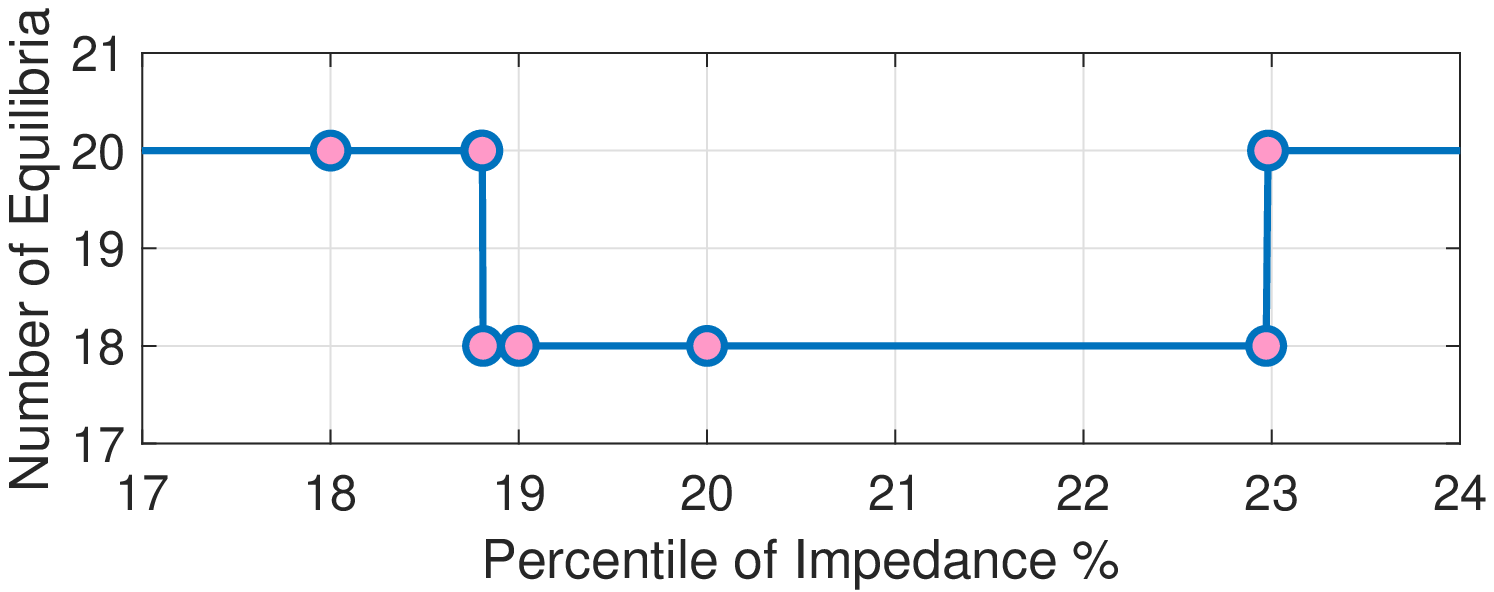}}
	\subfigure[Solution Pair Disappearance ]{\label{fig:case14type1+2}\includegraphics[width=0.48\columnwidth]{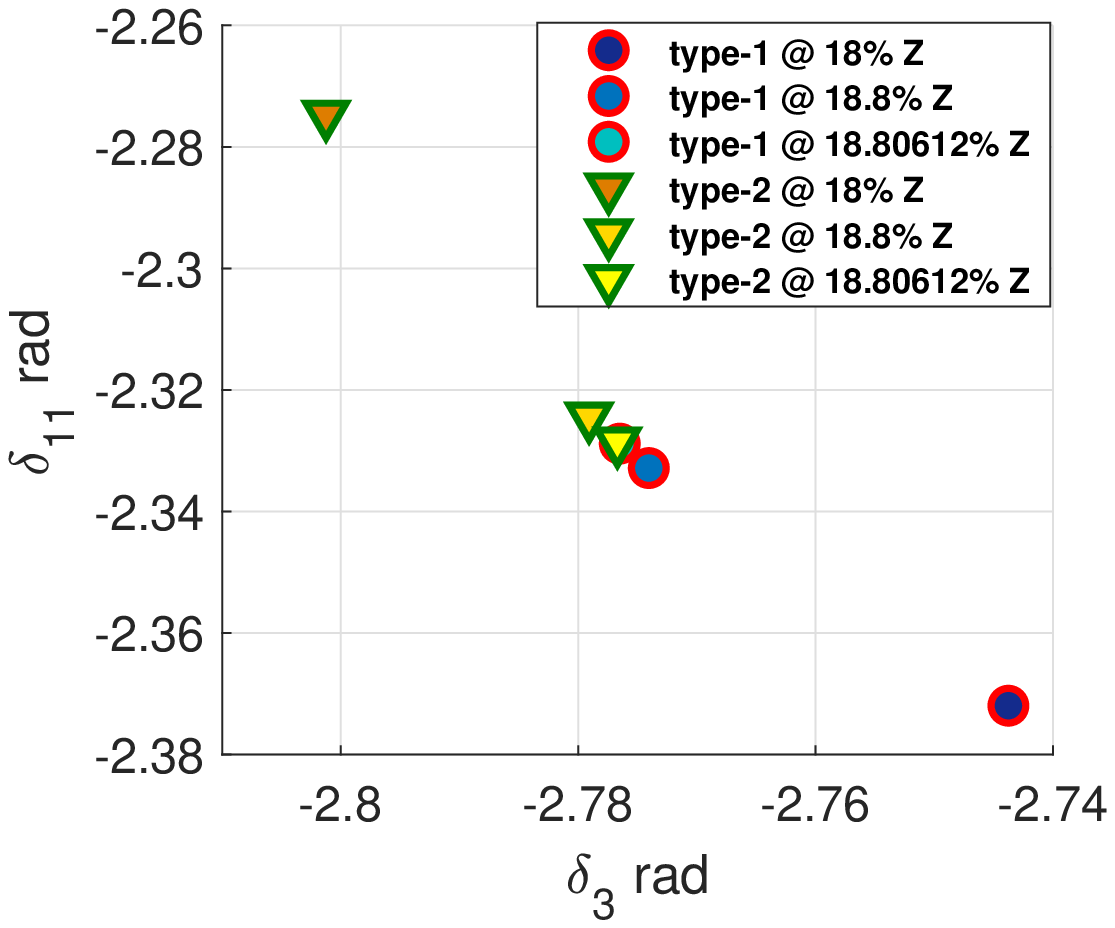}}
	\subfigure[Solution Pair Appearance ]{\label{fig:case14type0+1}\includegraphics[width=0.48\columnwidth]{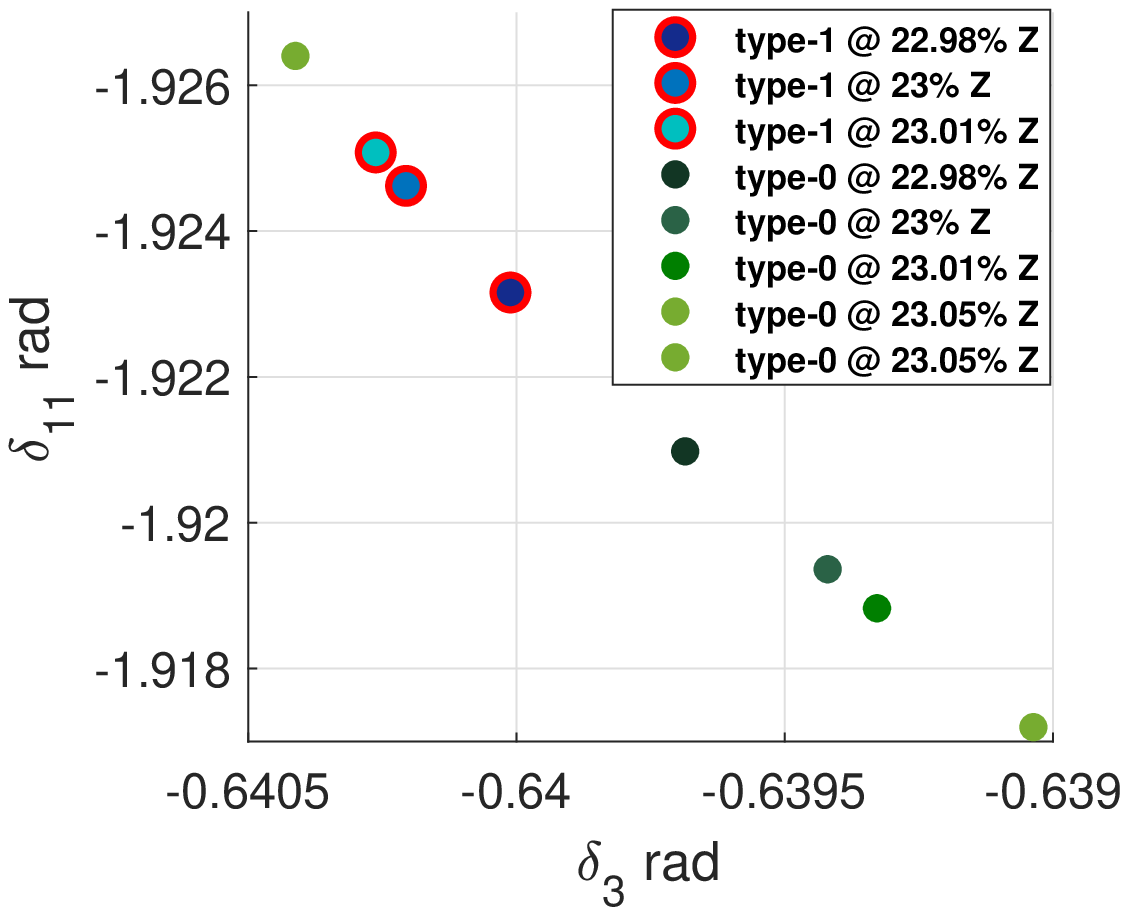}}
	\caption{Non-monotone Change of Solutions in 14-Bus Case} \label{fig:solu_change}
\end{figure}
The numbers of equilibria at different load models are summarized in Fig.~\ref{fig:num_equilibria}. Note that we apply \eqref{eq:ZP} as our load model for different proportions of constant power and impedance associated with $\alpha$. One observation is that a small proportion of power with a large proportion of impedance load induces much more equilibria than a large proportion of power with a small proportion of impedance load. This may be counter-intuitive because the constant impedance load is considered simpler than the constant power load. But as the constant power is gradually replaced by the constant impedance, it is equivalent to reducing the node power injection while increasing the shunt impedance. When the node power injection declines, it actually can intersect with more PV (QV) curves (surfaces), thus resulting in more equilibria. 

The second observation is that, as the load model approaches the constant impedance model, the number of equilibria increases much faster. This phenomenon may also be explained by the reason that many PV (QV) curves (surfaces) occur at a low node power injection level. It suggests that a small change of load model can result in many equilibria appearing or disappearing at a small constant power injection level. For example, in Fig.~\ref{fig:case14}, when the constant impedance increases from $80\%$ to $90\%$, the number of equilibria doubles. It may raise a reasonable concern about the validity of transient stability analysis based on some heuristically or conceptually assigned load models. However, this concern can be partially resolved by the observations in the next subsection when we further investigate the types of equilibria on the stable component of algebraic manifold.

Fig.~\ref{fig:case9} and Fig.~\ref{fig:case14zoom} show that the number of equilibria does not necessarily increase monotonically as the proportion of constant power decreases. In Fig.~\ref{fig:case9}, the number of equilibira declines around the pure constant impedance load model. In Fig.~\ref{fig:case14zoom}, the number of equilibia temporarily reduces at $18.81\%$ impedance proportion, and returns as the impedance proportion increases to $23\%$. Fig.~\ref{fig:case14type1+2} depicts the process of this temporary equilibria reduction. As the impedance proportion increases from $18\%$ to $18.80612\%$, a pair of equilibria collides on the real plane, and becomes a complex-valued solution pair. This collision happens between a type-1 equilibrium and a type-2 equilibrium. Fig.~\ref{fig:case14type0+1}, on the other hand, shows the acquisition of another equilibria pair which is not the same colliding equilibria pair in Fig.~\ref{fig:case14type1+2}. As the impedance proportion increases from $22.98\%$ to $23.01\%$, a pair of complex-valued solutions collides and becomes a real-valued equilibria pair from which one is type-0 and the other is type-1. At a particular proportion value between $23.01\%$ and $23.05\%$, the type-1 equilibrium turns into type-0. 

\subsection{Types of Equilibria}
\begin{figure}[!ht]
	\centering
	\subfigure[5-Bus Case]{\label{fig:case5type}\includegraphics[width=0.96\columnwidth]{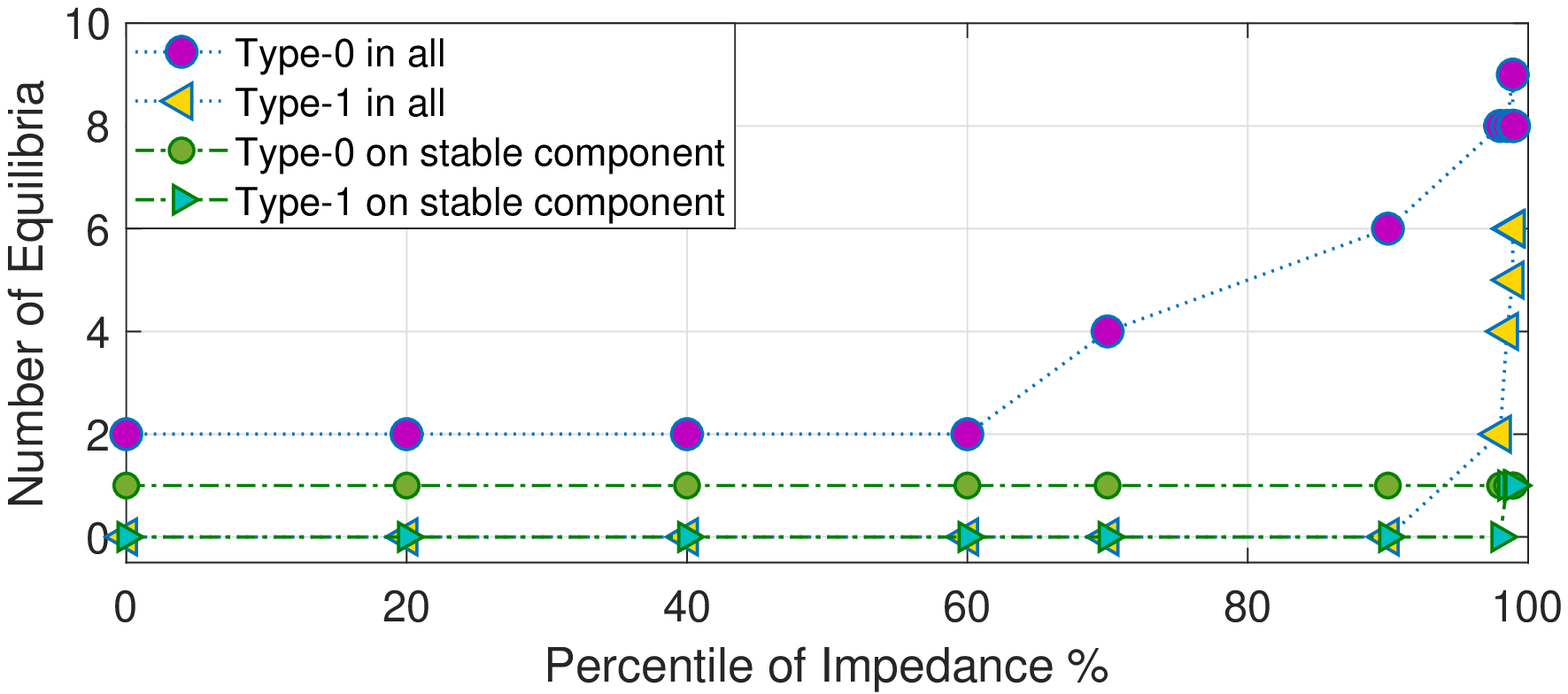}}
	\subfigure[9-Bus Case ]{\label{fig:case9type}\includegraphics[width=0.96\columnwidth]{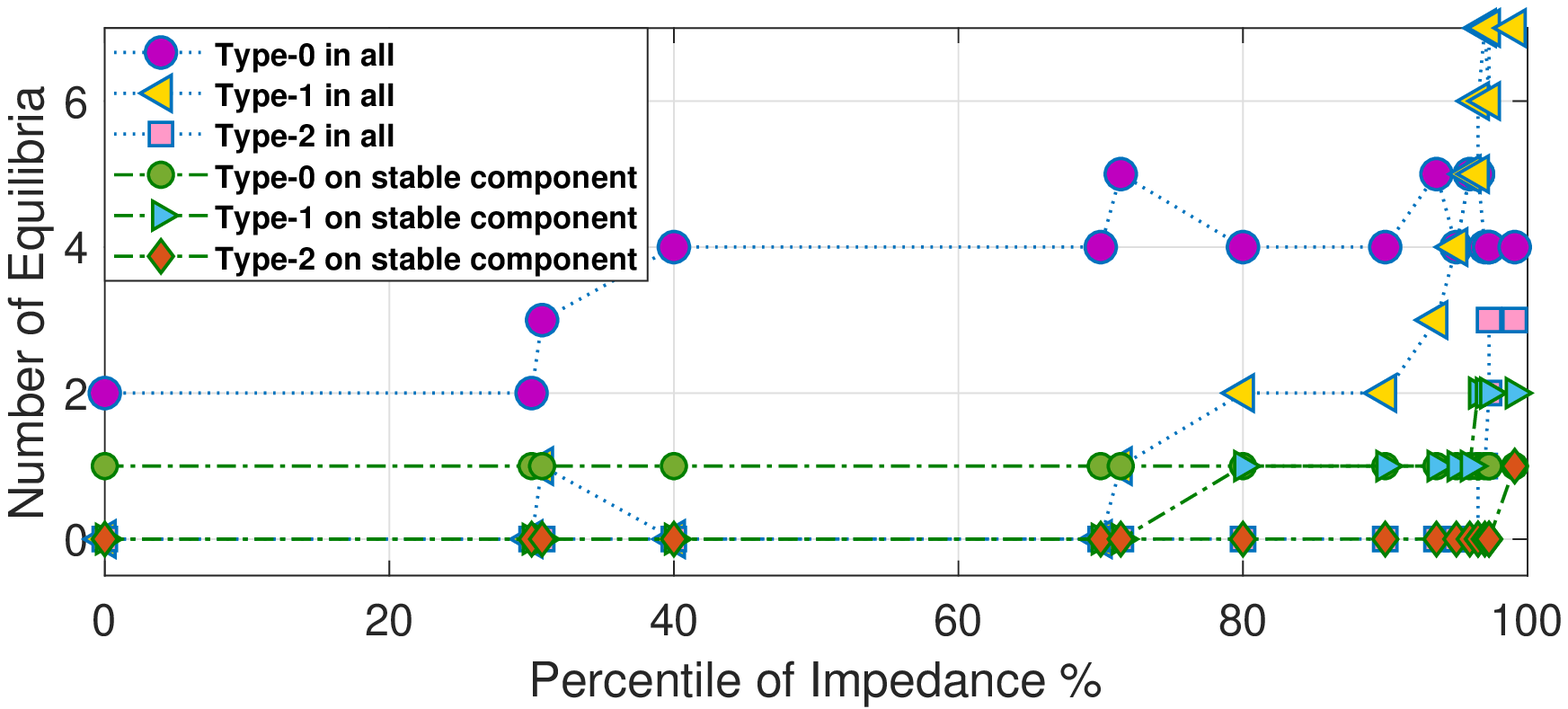}}
	\subfigure[14-Bus Case]{\label{fig:case14type}\includegraphics[width=0.96\columnwidth]{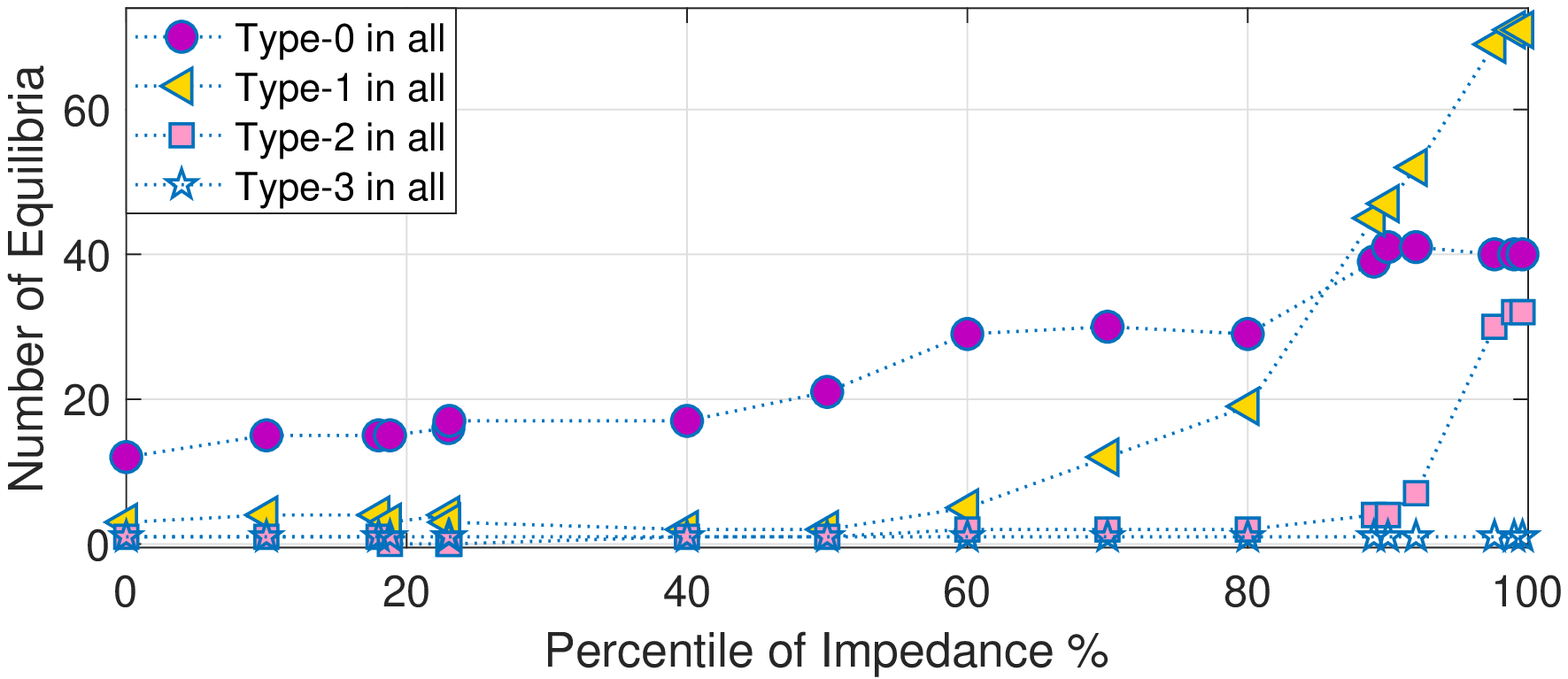}}
	\subfigure[14-Bus Case on Stable Manifold]{\label{fig:case14typestb}\includegraphics[width=0.96\columnwidth]{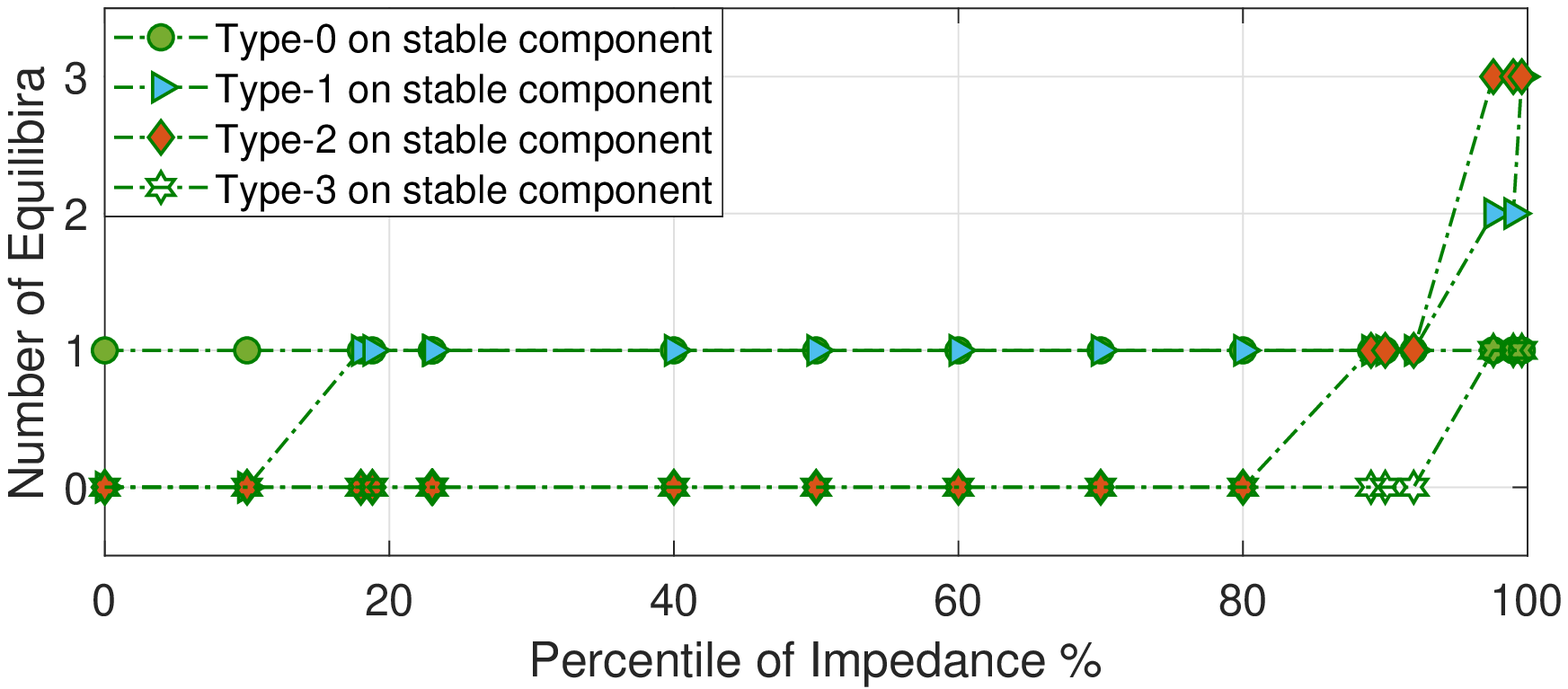}}
	\caption{Equilibria Types for Different Load Models} \label{fig:types}
\end{figure}
After enumerating equilibrium points for different load models, we evaluate their types displaying in Fig.~\ref{fig:types}. A DAE system \eqref{eq:DAE_COI} can be regarded as a confined dynamical system whose dynamical flows are restricted on the algebraic manifold defined by the algebraic equations of the DAE system. The type of an equilibrium point is determined by its differential part in \eqref{eq:Jx}. Therefore, multiple SEPs can exist on different components of the algebraic manifold. However, even some equilibria are classified as stable (type-0) in the analysis, the physical system may not be able to work at them. The algebraic part of a DAE system is usually derived from reducing fast dynamics, or other simplifications. Hence, an SEP in the DAE model may not be ``stable" in the physical sense. At this point, equilibria on the stable component of the algebraic manifold are of most interest because they are usually consistent with the types in the real situation. 

One observation from Fig.~\ref{fig:types} is that every tested case in each load model only exhibits one SEP (which is the high-voltage solution) on the stable component of the algebraic manifold. Ref. \cite{chiang1990:existence} discussed the uniqueness of load flow solution for radial distribution network, however, it is still an open question whether the SEP is unique or not on the stable component of algebraic manifold for a general power grid DAE system. Moreover, although the total number of equilibria can change drastically as shown in Fig.~\ref{fig:num_equilibria}, the number of equilibria on the stable component of algebraic manifold remains small for all tested cases in this paper. This observation may partially relieve the concern raised in the previous subsection: the change of load model can substantially alter the number of equilibria, but only very limited equilibria are modified on the stable component of algebraic manifold. More investigations are needed.

A more interesting finding is that the equilibria on the stable component of algebraic manifold become complicated with different types when the proportion of constant power reduces to a particular value, which is usually less than $20\%$. Fig.~\ref{fig:case9type} and Fig.~\ref{fig:case14typestb} present the increase of complexity for the equilibria. This phenomenon is somehow counter-intuitive because the constant power load is thought to be more complicated than the constant impedance load. Our analysis shows the opposite, at least in the sense that a small proportion of constant power with a large proportion of constant impedance will result in more complex dynamical behaviors.

\begin{table}[!ht]
	\centering
	\caption{Selected Equilibria from 9-Bus Case}
	\begin{tabular}{c|c|c|c|c|c}
		\hline 
		\hline
		Bus        & $V_1$     & $V_2$     & $V_3$     & $V_4$                            & $V_5$                            \\ \hline
		Equilibrium 1 & 0.3519 & 0.7600 & 0.7565 & { 0.0696} & { 0.1584} \\ \hline
		Equilibrium 2 & 0.5889 & 0.8282 & 0.7490 & 0.4042                        & { 0.0037} \\ \hline \hline
		Bus        & $V_6$     & $V_7$     & $V_8$     & $V_9$                            & Type                          \\ \hline
		Equilibrium 1 & 0.6060 & 0.5874 & 0.5939 & { 0.1488} &  1     \\ \hline
		Equilibrium 2 & 0.5948 & 0.6505 & 0.7081 & 0.4883                        & 0                             \\ \hline \hline
	\end{tabular} \label{table:case9}
\end{table}
Another interesting discovery may also counter people's intuition: neither a type-1 equilibrium point (on the stable component of the algebraic manifold) should admit only one low bus voltage, nor a solution with only one low bus voltage should be type-1. For example, in Table~\ref{table:case9}, we present two selected equilibria from the 9-bus case at $97\%$ proportion of impedance load. The first equilibrium is type-1 on the stable component of algebraic manifold. The voltage magnitudes on bus 4, 5 and 9 are all below $0.2$ p.u. While the second equilibrium is type-0 on the unstable component of algebraic manifold. It only has one voltage magnitude at bus 5 below $0.2$ p.u. Other tested cases also exhibit a similar phenomenon. This observation suggests that an initial guess with only one low voltage magnitude cannot necessarily lead the Newton's method\footnote{Any method that requires a starting point may fail to identify a type-1 equilibrium point from the mentioned initial guess.} to find a type-1 equilibrium point. It also suggests that an equilibrium with only one low voltage magnitude can also be another type.

\section{Stability Regions with Different Load Models}
   	\label{sec:stability}
   	Section III shows that power system DAEs can have multiple SEPs on different components of the algebraic manifold. This section will investigate the load model influence on the stability region (or the stability boundary which is the boundary surface of region of attraction) with respect to the high-voltage SEP. Unlike the power system ODE dynamical models whose stability boundaries are composed of the stable manifolds of certain type-1 UEPs, the unstable manifolds of which converge to the SEP, the stability boundaries of power system DAE dynamical models consist of two components: the first component is the same as the ODE model, while the second one is a set of points whose trajectories reach singular surfaces \cite{venkatasubramanian1992:stability,venkatasubramanian1995:dynamics,chiang2015,wu2006:geometry}.

\begin{figure*}
	\centering
	\subfigure[$0\%$ Impedance]{\label{fig:case9_00_tsb}\includegraphics[width=0.62\columnwidth]{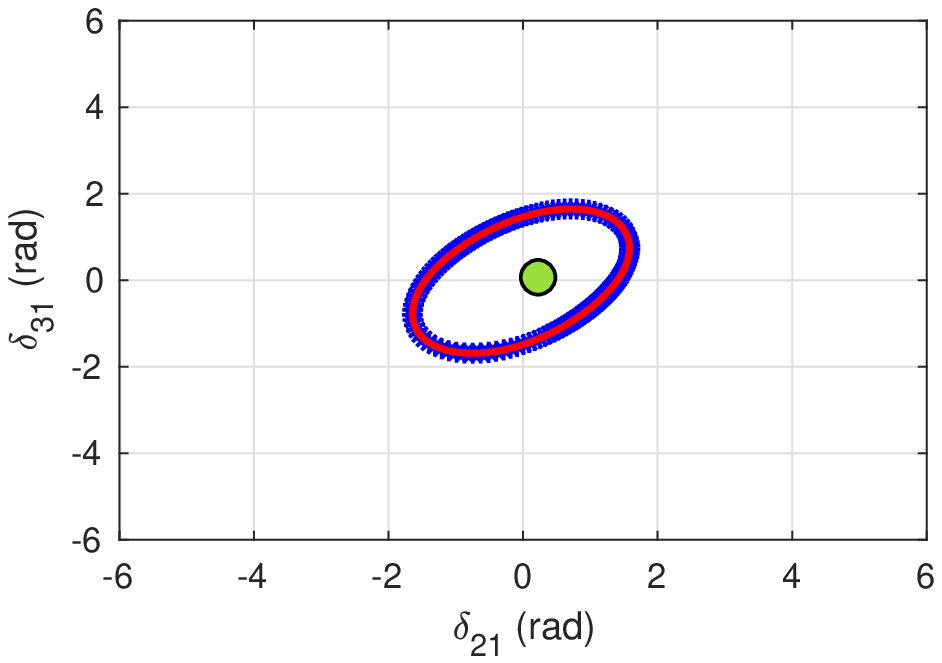}}
	\subfigure[$40\%$ Impedance ]{\label{fig:case9_04_tsb}\includegraphics[width=0.62\columnwidth]{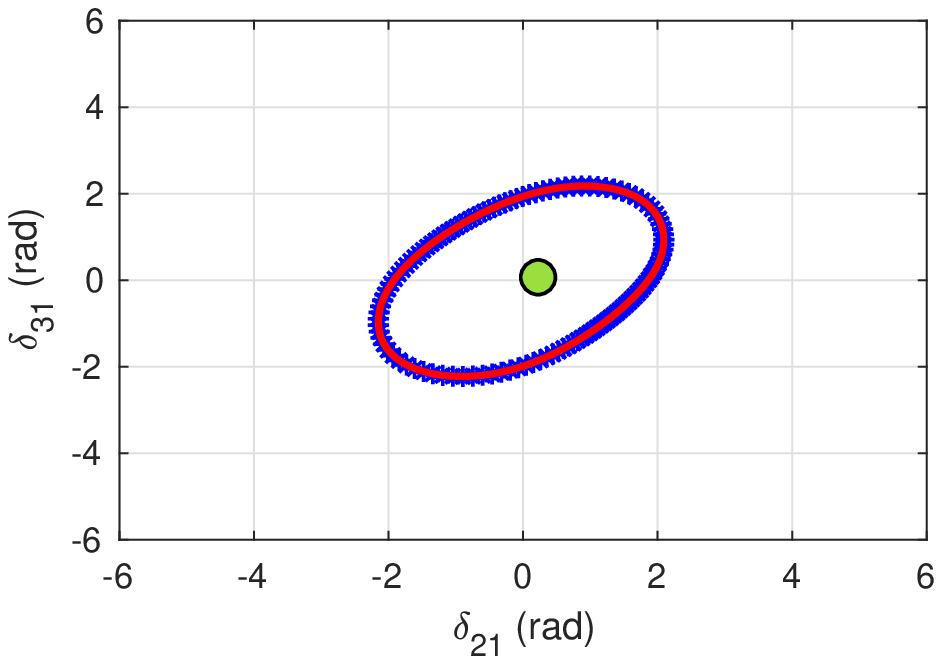}}
	\subfigure[$80\%$ Impedance]{\label{fig:case9_08_tsb}\includegraphics[width=0.62\columnwidth]{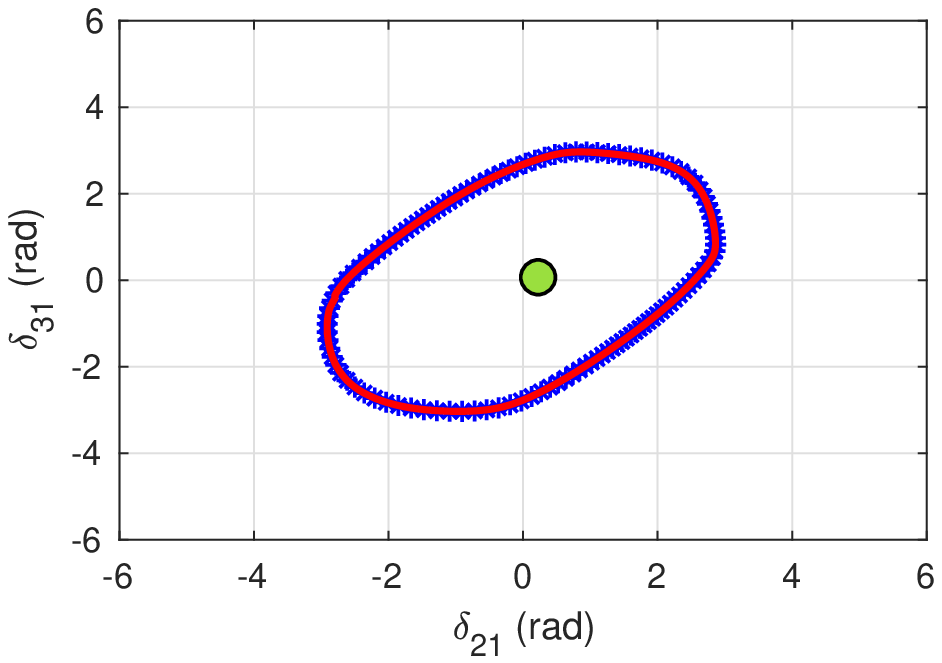}}\\
	\subfigure[$90\%$ Impedance]{\label{fig:case9_09_tsb}\includegraphics[width=0.62\columnwidth]{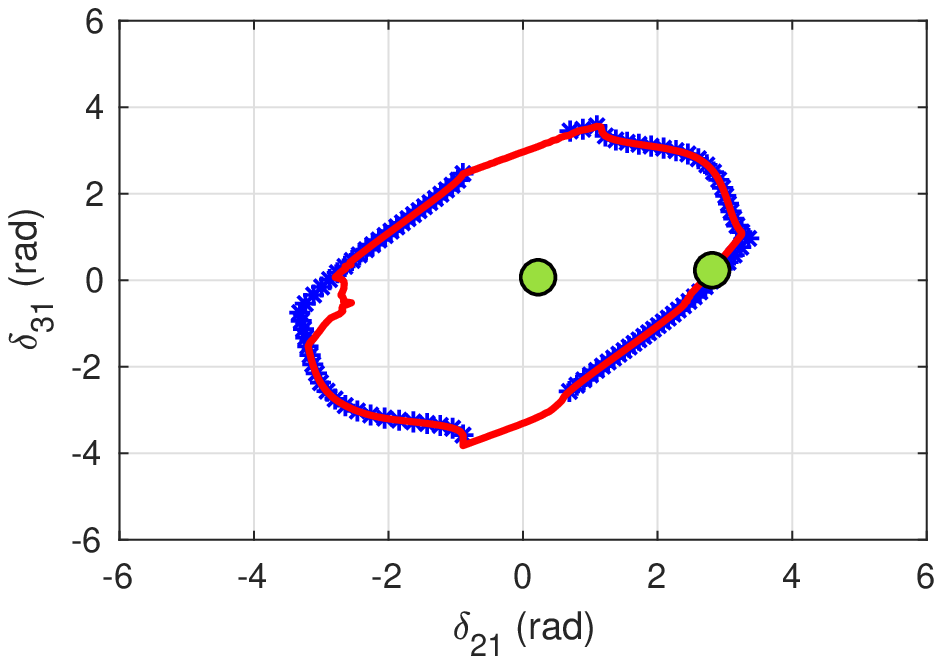}}
	\subfigure[$95\%$ Impedance ]{\label{fig:case9_095_tsb}\includegraphics[width=0.62\columnwidth]{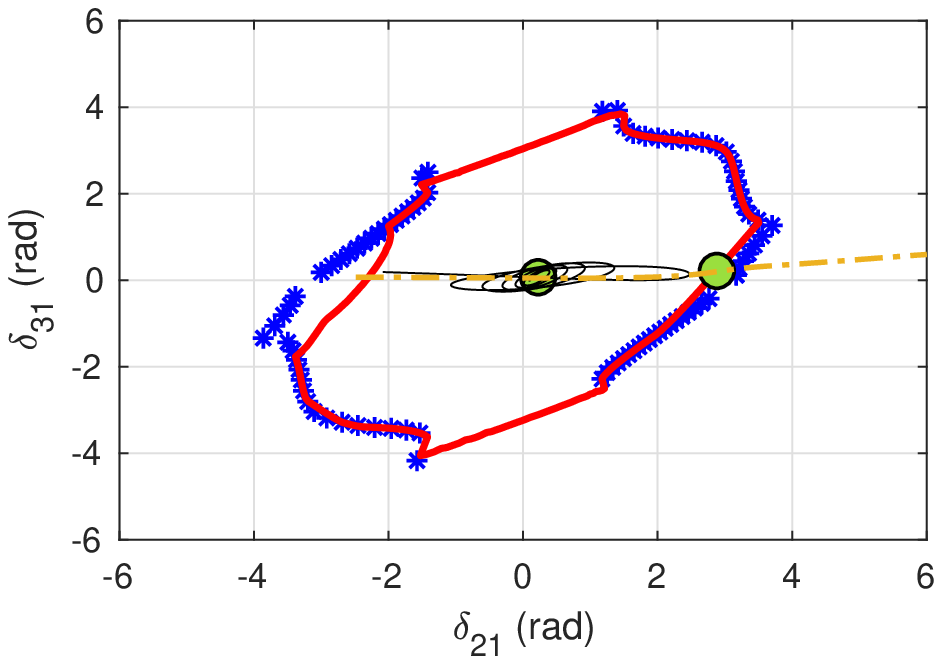}}
	\subfigure[$100\%$ Impedance]{\label{fig:case9_110_tsb}\includegraphics[width=0.62\columnwidth]{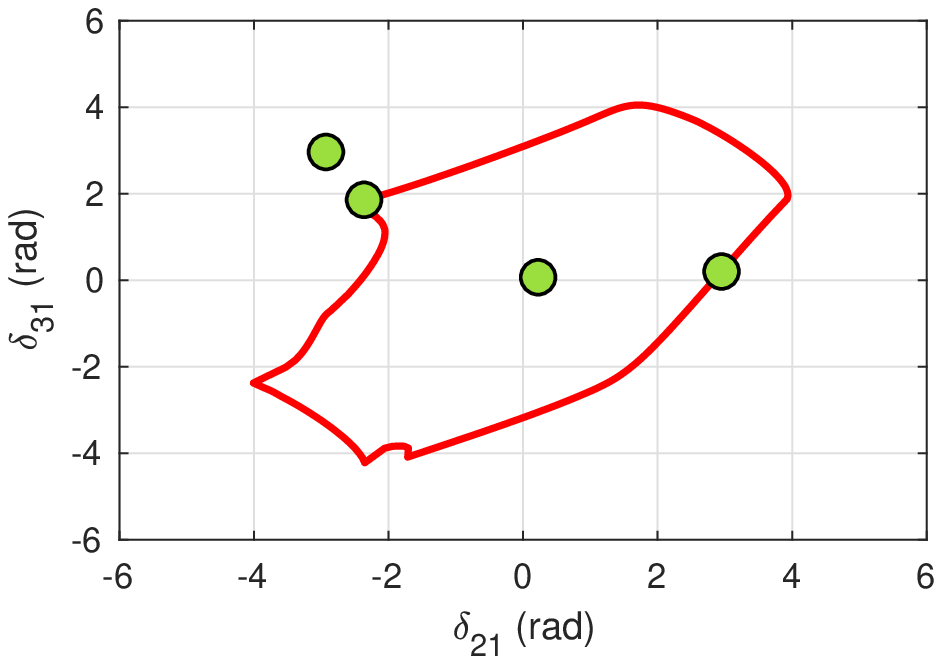}}
	\caption{9-Bus Case Stability Boundary} \label{fig:case9_tsb}
\end{figure*}

Fig. \ref{fig:case9_tsb} shows the cross-sections of the stability regions on the zero-speed plane\footnote{Note that in our COI framework, the stable equilibrium does not need to be at the rated speed. It only converges to the speed of COI.} for the 9-bus case with different load models. These intersections can be regarded as particular slices of the corresponding stability regions of the power system DAE models in the $\delta-\omega$ space. For simplicity, we refer this intersection to ``stability region" in the following discussions. In Fig. \ref{fig:case9_tsb}, green solid circles are equilibria on the stable component, red curves represent stability boundaries identified by Algorithm \ref{alg:tsb}, and blue stars represent the projection of singular surfaces on the relative angle plane.

As shown in Fig. \ref{fig:case9_00_tsb}-\ref{fig:case9_08_tsb}, when the proportion of impedance load increases from $0\%$ to $80\%$, the stability region of the 9-bus case expands gradually. During this process, there is only one equilibrium, which is stable, and the stability boundary is exactly the same as the singularity boundary.

When the proportion of impedance load reaches $90\%$, the singularity boundaries break up into multiple segments in the plot, and a type-1 UEP appears around $(\delta_{21},\delta_{31}) = (2.4,0.2)$, whose stable manifolds contribute to a small portion of the stability boundaries around this UEP, as shown in Fig. \ref{fig:case9_09_tsb} and \ref{fig:case9_095_tsb}. Note that the stability boundary near $(-3,-1)$ does not overlap with any singularity boundaries because the stability region shown in the plot is just an angle slice that does not necessarily comply with the angle slice of the singularity boundary. 
The dash dot line in Fig. \ref{fig:case9_095_tsb} represents an unstable trajectory initialized at $(\delta_{21},\delta_{31}) = (-2.08,0.19)$, i.e. a point outside stability region but inside the singularity region. 
As a comparison, the black trajectory initialized at $(\delta_{21},\delta_{31}) = (-2.48,0.07)$, i.e. a point in the stability region but close to the boundary, first crosses the SEP and travels near the type-1 UEP. But it converges to the SEP eventually.

When the load is represented by a pure impedance, the singularity boundary completely disappears and the stability boundary is fully determined by stable manifolds of the two type-1 UEPs, as shown in Fig. \ref{fig:case9_110_tsb}. 

Another interesting finding is that except for the case with $100\%$ impedance load model, the majority of the stability boundary is determined by the singular surface. When the proportion of constant power reduces from $100\%$ to less than $5\%$, the singular surface expands in the relative angle subspace and disappears eventually. In the meantime, the expanded area tends to include more UEPs, which contribute to a portion of the stability boundary.

\begin{algorithm}[tbhp]
	\caption{Numerical identification of the stability boundary}
	\label{alg:tsb}
	\begin{algorithmic}[1]
		\State{In the $\delta_{21}-\delta_{31}$ plane, set up $M$ unit vectors with angles respectively taking $0$, $360/M$, ..., $360(M-1)/M$, say $\textbf{n}_{1}$, ...,  $\textbf{n}_{M}$. $M=180$ is used in this paper.}
		
		\State{Set $r$ to be a small step, e.g. 0.1 used in this paper. Along the direction determined by each of the unit vectors, say $\textbf{n}_{j}$, let $(\delta_{210},\delta_{310}) = r\textbf{n}_{j}$ and conduct steps 3-6 below.}
		
		\State{Use $(\tilde{\delta}_{1},\tilde{\omega}_{1},\tilde{\delta}_{2},\tilde{\omega}_{2},\tilde{\delta}_{3},\tilde{\omega}_{3}) = (0,0,\delta_{210},0,\delta_{310},0)$ as the initial state to numerically solve DAEs in (\ref{eq:DAE_COI}) over a period of time (5 seconds in this paper) for $\delta_{21}(t)$ and $\delta_{31}(t)$.}
		
		\State{If $r<\epsilon$ ($\epsilon=0.001$ in this paper) record $(\delta_{210},\delta_{310})$ as an estimate of the stability boundary in direction $\textbf{n}_{j}$ and then go to step 2 for next direction.}

        \State{If $\text{max}(\delta_{21})-\text{min}(\delta_{21})>2\pi$ or $\text{max}(\delta_{31})-\text{min}(\delta_{31})>2\pi$, let $r = r/2$, $(\delta_{210},\delta_{310}) = (\delta_{210},\delta_{310}) - r\textbf{n}_{j}$ and go to step 3.}
        
        \State{Let $(\delta_{210},\delta_{310}) = (\delta_{210},\delta_{310}) + r\textbf{n}_{j}$ and go to step 3.}
	\end{algorithmic}
\end{algorithm}

\section{Algebraic Manifolds with Different Load Models}
	\label{sec:manifold}
	\begin{figure*}
	\centering
	\subfigure[$0\%$ Impedance]{\label{fig:case9_00}\includegraphics[width=0.65\columnwidth]{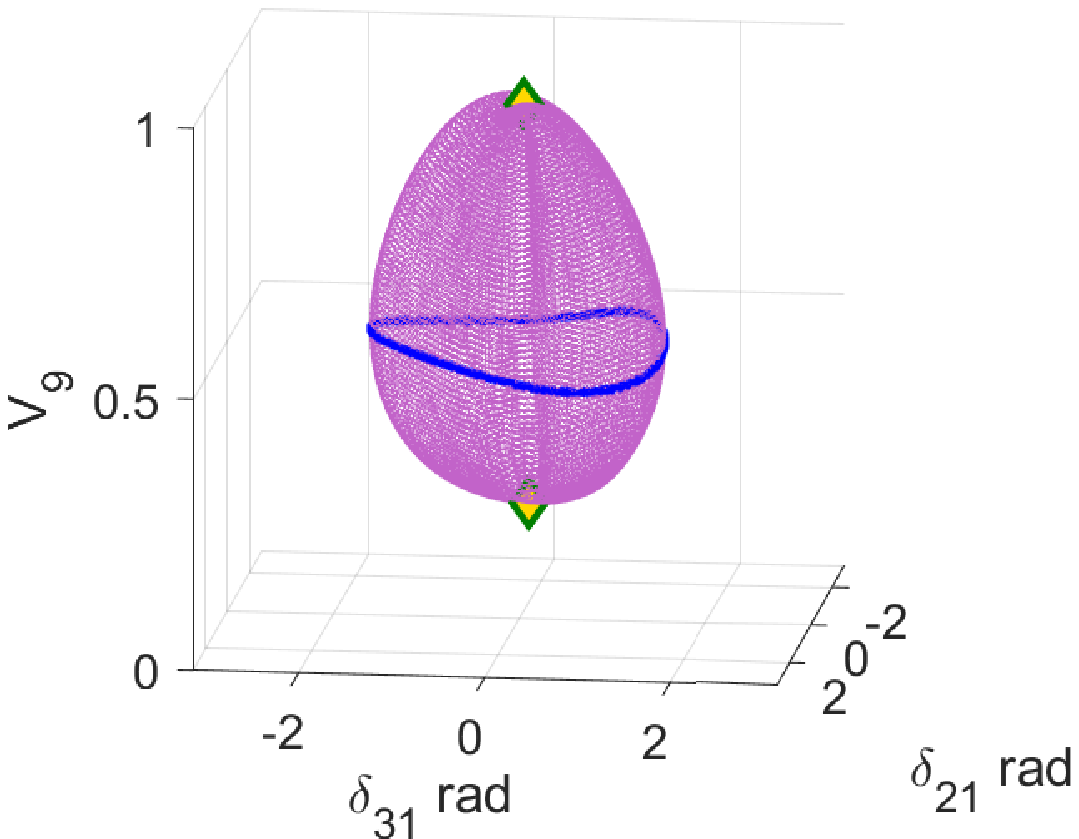}}
	\subfigure[$20\%$ Impedance ]{\label{fig:case9_02}\includegraphics[width=0.65\columnwidth]{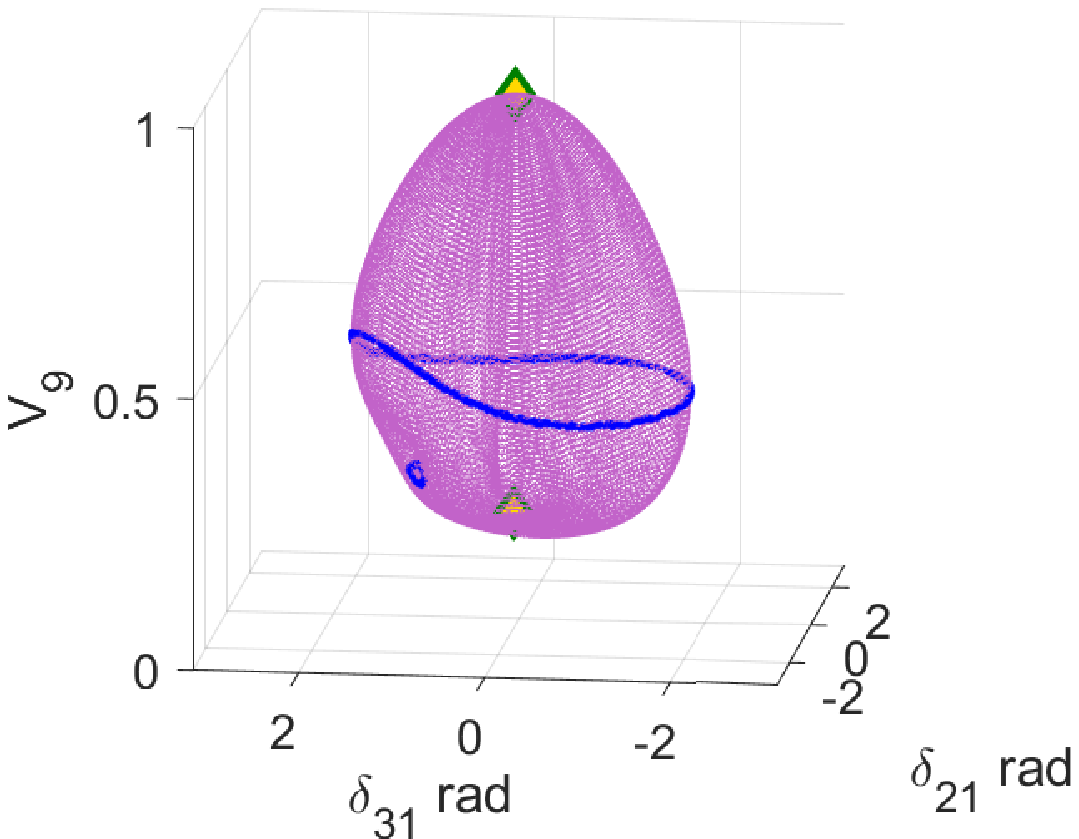}}
	\subfigure[$40\%$ Impedance]{\label{fig:case9_04}\includegraphics[width=0.65\columnwidth]{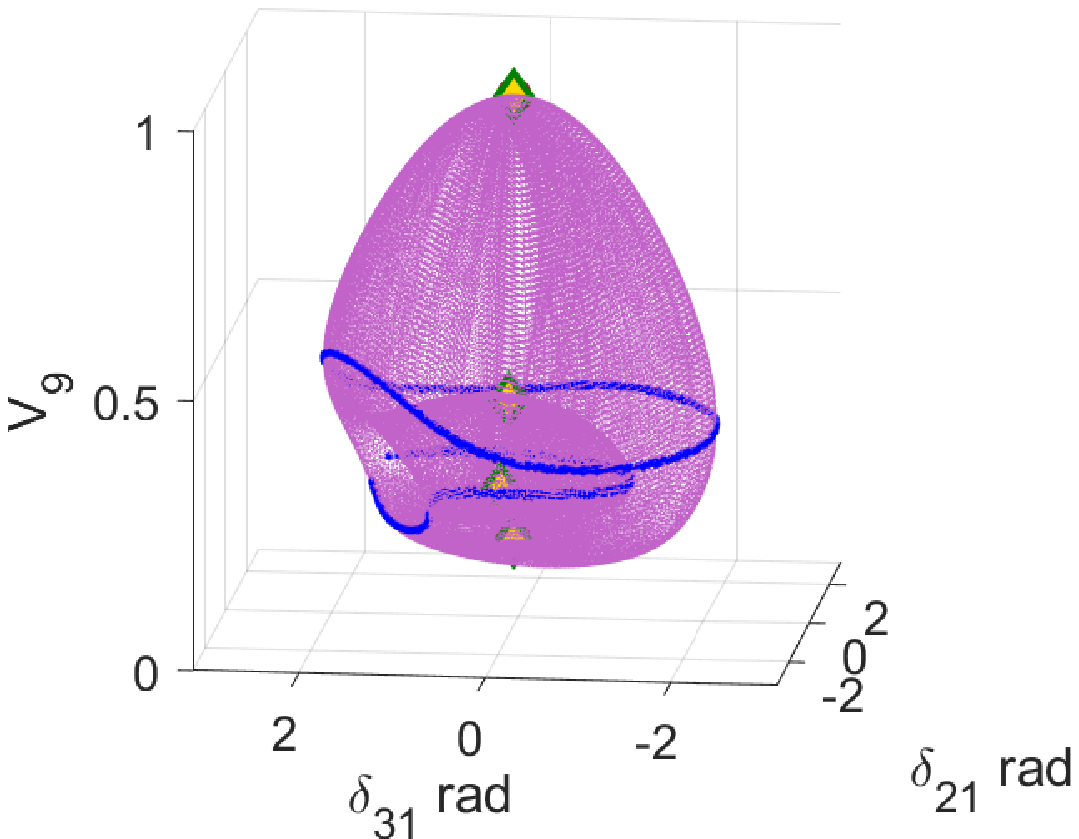}}\\
	\subfigure[$60\%$ Impedance]{\label{fig:case9_06}\includegraphics[width=0.65\columnwidth]{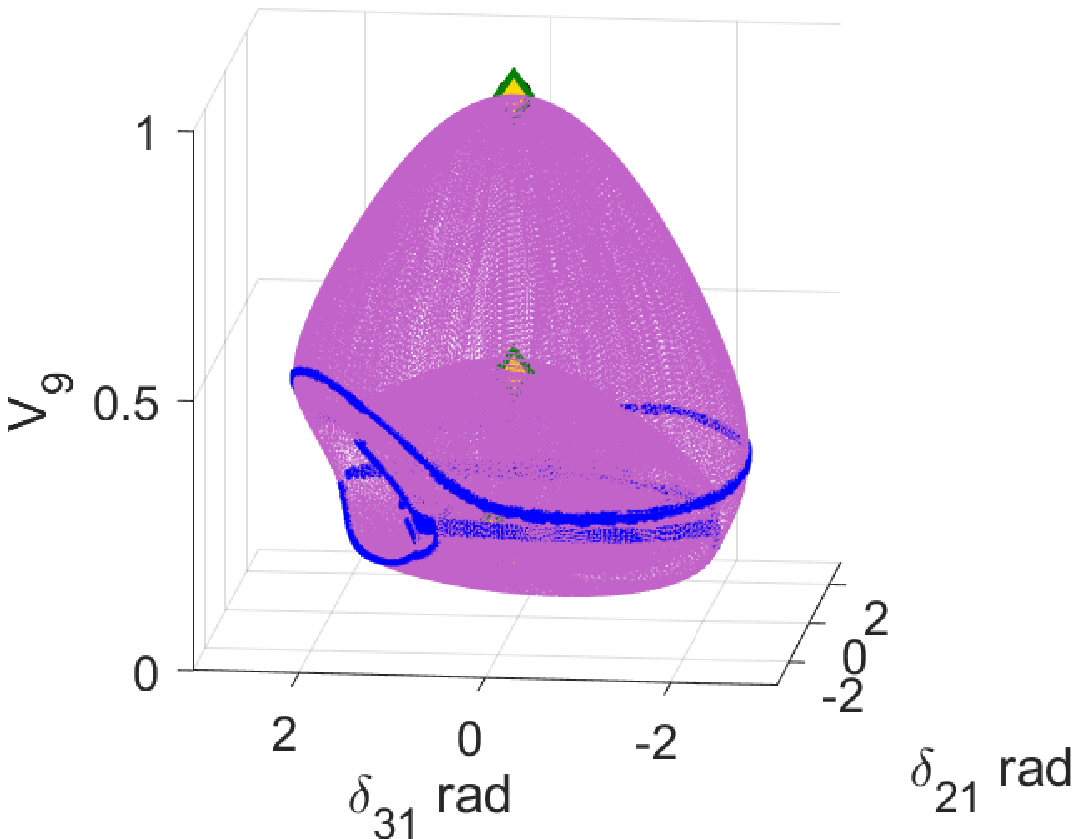}}
	\subfigure[$80\%$ Impedance ]{\label{fig:case9_08}\includegraphics[width=0.65\columnwidth]{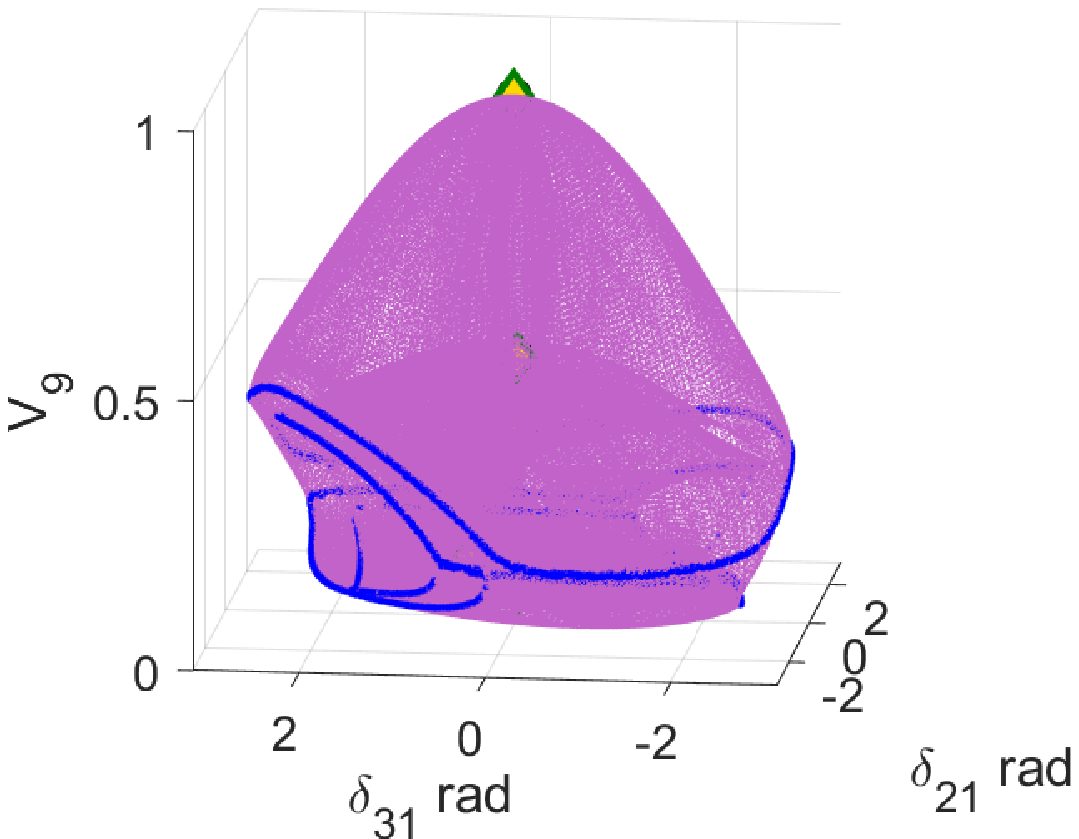}}
	\subfigure[$93.1\%$ Impedance]{\label{fig:case9_0931}\includegraphics[width=0.65\columnwidth]{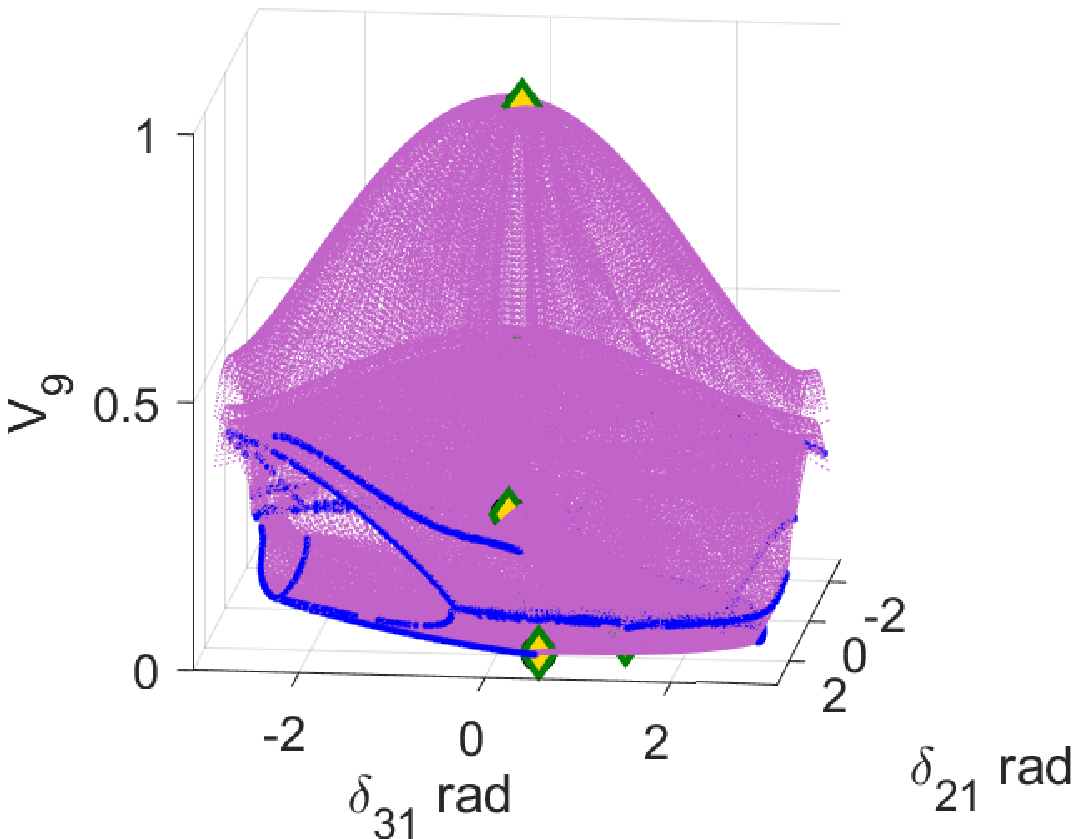}}
	\caption{9-Bus Case Algebraic Manifolds} \label{fig:case9_mnfld}
\end{figure*}
Our analysis started with equilibria in Section~\ref{sec:equilibria}, and extended to stability regions on the stable component of algebraic manifold in Section~\ref{sec:stability}. Now in this section we directly investigate the algebraic manifold to acquire a global view of how the load model influences the DAE system. The algebraic manifold is defined by the algebraic set of \eqref{eq:hV}, \eqref{eq:hPd}, and \eqref{eq:hQd}. We only focus on the 9-bus case because it is complicated enough to demonstrate interesting phenomena, while not computationally extensive. 

Fig.~\ref{fig:case9_mnfld} depicts algebraic manifolds under different load models. Specifically, the $X$-axis represents the generator angle $\tilde{\delta}_2$ which equals to the angle difference $\delta_{21} \mydef \delta_2-\delta_1$ as described by \eqref{eq:delative_del}. The $Y$-axis similarly represents $\tilde{\delta}_3=\delta_{31}$. The $Z$-axis is chosen to be the voltage magnitude on bus-$9$. The algebraic manifolds are shown in pink surfaces on which the blue curves represent the singularity boundary surfaces. The yellow diamonds are the equilibria that we analyzed in Section~\ref{sec:equilibria}.

From Fig.~\ref{fig:case9_00} to Fig.~\ref{fig:case9_0931} one can observe that the algebraic manifold enlarges as the proportion of impedance load increases. This is consistent with the finding that constant power model induces smaller stability region than the constant impedance model \cite{2009Sorrentino}. Since the $X$ and $Y$ axes represent angles, they should repeat themselves by any $2n \pi$. When the impedance model dominates the load model, the algebraic manifold connects to its $2n \pi$ replica, as shown in Fig.~\ref{fig:case9_0931}. It suggests that the DAE system dynamics may stabilize after several angular cycles of $2\pi$. 

\begin{figure}
	\centering
	\subfigure[$40\%$ Impedance]{\label{fig:case9_04_inner}\includegraphics[width=0.9\columnwidth]{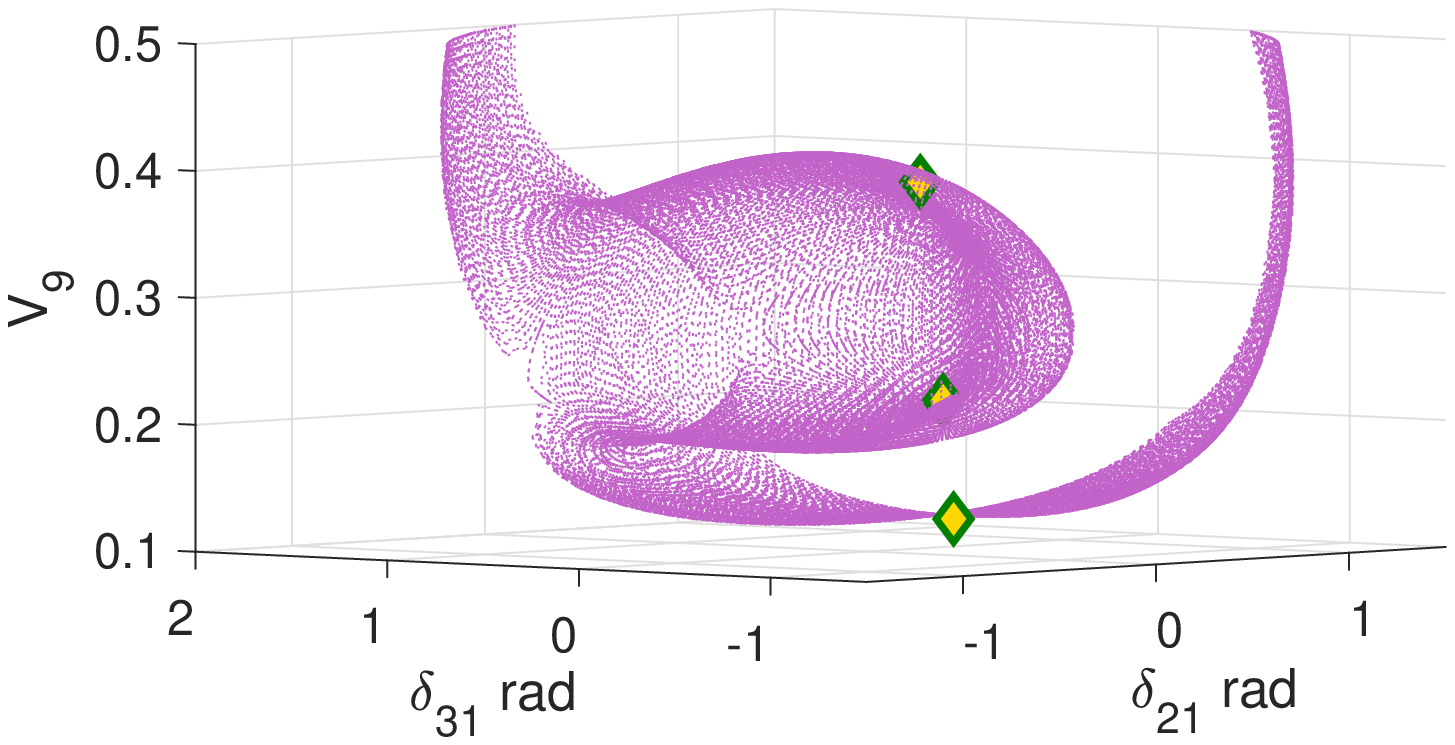}}
	\subfigure[$60\%$ Impedance]{\label{fig:case9_06_inner}\includegraphics[width=0.9\columnwidth]{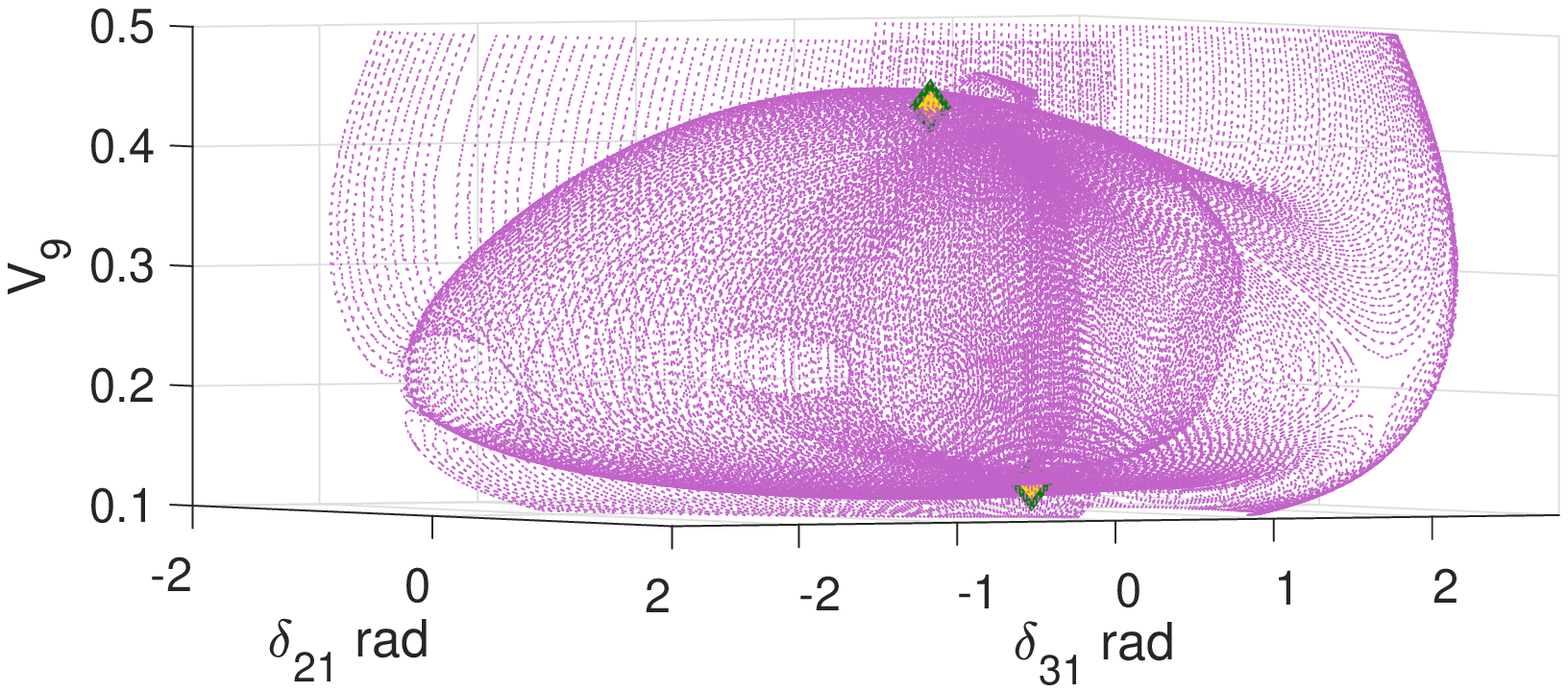}}
	\caption{9-Bus Case Algebraic Manifolds: Inner Bubbles} \label{fig:case9_mnfld_inner}
\end{figure}
A new finding is that a large proportion of impedance in the load really complicates the geometry of the algebraic manifold, and substantially alters the topology of the manifold as well. For example, with constant power load model, the algebraic manifold depicted in Fig.~\ref{fig:case9_00} is homeomorphic to a sphere. When the impedance load increases to $20\%$ in Fig.~\ref{fig:case9_02}, a small depression area appears. It induces negative curvatures and creates its own singularity boundary which is shown as the small blue circle in the plot. When the impedance proportion increases further to $40\%$, the small depression area becomes a concave bubble inside the outer manifold and introduces more equilibria on it, shown in Fig.~\ref{fig:case9_04} (global view) and Fig.~\ref{fig:case9_04_inner} (local view). As we increase the load impedance to $60\%$, two inner bubbles reach the other side of the outer surface, punctuating the manifold to create two tunnels shown in Fig.~\ref{fig:case9_06} (global view) and Fig.~\ref{fig:case9_06_inner} (local view). Now the manifold is no longer homeomorphic to a sphere, but homeomorphic to two tori glued together. Increasing the impedance proportion further enlarges the tunnels and creates more complicated structures as shown in Fig.~\ref{fig:case9_08} and \ref{fig:case9_0931}. 

In this particular example, however, both the geometric and the topological changes occur on the unstable components of algebraic manifold. Whether the stable component can exhibit a similar change requires further investigations. 

\section{Conclusion} 
   	\label{sec:concl}
   	Load models are recognized being influential on voltage behaviors and transient dynamics. Numerous works have been done on this topic to build appropriate load models and to better capture static and dynamical behaviors.

In this paper, we specifically focused on the geometric and topological changes of the solution sets for a power system DAE model with different load models. A few findings were recorded and analyzed when the load model continuously transitions from the constant power model to the constant impedance model. One major counter-intuitive discovery is that a load model with a large proportion of constant impedance and a small proportion of constant power introduces much more complex geometric and topological structures than a load model with a small proportion of constant impedance and a large proportion of constant power. Specifically, the number of equilibria increases dramatically when the constant power is largely replaced by the constant impedance. However, this increase of equilibrium quantity is not necessarily monotone. We also emphasized that the appearance of type-1 UEP is not necessarily associated to a single node experiencing ultra-low voltage magnitude. The stability region, as expected, starts from a small region determined by the singular surface, expands when the proportion of constant impedance increases, and is finally constrained by the stable manifolds of type-1 UEPs. In this case, the algebraic manifold is connected to all its modulo $2 \pi$ replica. The fundamental group of the algebraic manifold is, however, more complicated when the load is dominated by the impedance. 

These findings certainly provide deep insights to the fundamental geometric and topological structures of power system dynamics associated with different load models. Future investigations will include, not limited to, how to identify changes of equilibria only on the stable component of algebraic manifold; what changes of other load models can exert on power system dynamical models; how to identify the appropriate distance from the SEP to the nearest stability boundary point in the power system DAE model.

\section*{Acknowledgement} 
The authors gratefully acknowledge Prof. Bernard Lesieutre, Prof. Chris Demarco, Prof. Franz-Eric Wolter, and Prof. Marija Ilic for many helpful discussions. We also thank Prof. Eytan Modiano and Prof. Le Xie for their great supports.

\bibliographystyle{ieeetr}
\bibliography{References}

\appendices
\section{9-Bus System Parameters} 
\label{sec:Appendix I}
The dynamical parameters are selected from the Appendices of \cite{anderson2008:power} and have been converted to the same base power at $100$ MVA.
\begin{table}[!ht]
	\centering
	\caption{5-Bus Case Dynamical Data}
\begin{tabular}{c|c|c|c|c|c}
\hline
Bus & Rated Power & $X_d'$ & r      & $T_{do}'$     & H                             \\ \hline
1   & 100 MVA     & 0.2200                  & 0.0035 & 5.900 & 4.9850 \\ \hline
5   & 147.1 MVA   & 0.2033                  & 0.0023 & 4.300                        & 4.3100 \\ \hline 
\end{tabular}\label{table:5-bus}
\end{table}

\begin{table}[!ht]
	\centering
	\caption{9-Bus Case Dynamical Data}
	\begin{tabular}{c|c|c|c|c|c}
		\hline
		Bus & Rated Power & $X_d'$     & r      & $T_{do}'$                           & H                              \\ \hline
		1   & 233 MVA     & 0.1391 & 0.0007 &  5.1400 & 9.5996  \\ \hline
		2   & 270 MVA     & 0.0948 & 0.0006 & 4.8000                        & 11.1510 \\ \hline
		3   & 270 MVA     & 0.0948 & 0.0006 & 4.8000                        & 11.1510                        \\ \hline
	\end{tabular}\label{table:9-bus}
\end{table}

\begin{table}[!ht]
	\centering
	\caption{14-Bus Case Dynamical Data}
	\begin{tabular}{c|c|c|c|c|c}
		\hline
		Bus & Rated Power & $X_d'$     & r      & $T_{do}'$                           & H                             \\ \hline
		1   & 330 MVA     & 0.0961 & 0.0005 &  6.0000 & 9.900 \\ \hline
		2   & 147.1 MVA   & 0.2033 & 0.0023 & 4.3000 & 4.3100\\ \hline
		3   & 192 MVA     & 0.1641 & 0.0019 & 5.0000                        & 5.9520                        \\ \hline
		6   & 100 MVA     & 0.2200 & 0.0035 & 5.9000                        & 4.9850                        \\ \hline
		8   & 100.1 MVA   & 0.3137 & 0.0049 & 6.5500                        & 3.1201                        \\ \hline
	\end{tabular}\label{table:14-bus}
\end{table}

\end{document}